# Significant and megathrust earthquake predictions by real-time monitoring of the genesis processes with Physical Wavelets

Fumihide Takeda [1,2,*]

[1] Takeda Engineering Consultant Co., Hiroshima, Japan
[2] Earthquake Prediction Institute, Imabari, Japan

[*] f_takeda@tec21.jp

Physical Wavelets can offer real-time significant and megathrust earthquake predictions and disaster prevention warnings up to three months in advance, saving lives and minimizing damages.

## Contents

**Abstract**

Japan's seismograph and Global Positioning System (GPS) networks provide a detailed record of subduction-zone earthquakes (EQs) in the form of stochastic and irregular time series data. Physical Wavelets are deterministic mathematical operators used for analyzing these time series data to define the equations of significant and megathrust EQ genesis processes. These equations can be used to predict EQ hypocenters (focuses), fault movements, sizes, and rupture times in real-time with accuracies within a day before the EQ occurrences up to three months. The information can be used to issue disaster prevention warnings like those used for typhoons. For instance, Physical Wavelets analyses of the 2011 Tohoku M9 EQ illustrated a megathrust EQ genesis process of fifteen months and a tsunami genesis process of the last three months, which could have provided real-time disaster prevention warnings and hazard mitigation measures leading up to the event.

**1 Introduction**

The Earth's lithosphere consists of three layers: the brittle upper crust (B), the ductile lower crust (D), and the ductile uppermost mantle. Plate-driving forces generate steady-state creep in the ductile layers, accumulating stress in the brittle upper crust (B) through the D-B transition region, which couples the three layers [1]. High ductile strain rates can generate earthquakes (EQs) of varying sizes in the B, resulting in stochastic and complex EQ activity. The stress state is a frictional failure and the principal stress components are vertical and horizontal to the Earth's surface [1]. A self-organizing system [2] or a self-organized criticality (SOC) hypothesis [3] has been proposed to explain the complexity. However, these hypotheses cannot explain the genesis process of the 2011 Tohoku M9 EQ observed with Japanese seismic and GPS networks [4–8]. A subtle depth-scale dependence of the Tohoku EQ originating from the D-B transition region [7] contradicts the expected scale-invariance (self-similarity) [9, 10] and the SOC hypothesis [3].

To resolve these issues and significantly improve the current probabilistic megathrust EQ and tsunami hazard mitigation [11], we updated a Japanese patent [6] based on two observations – seismic [7] and GPS [8]. Physical Wavelets (PWs), deterministic mathematical operators, were used to quantify a megathrust EQ genesis process of fifteen months and a tsunami genesis process of the last three months using the 2011 Tohoku EQ GPS record. The GPS records are non-differentiable daily displacement time series data collected at the GPS stations in the Tohoku subduction zone.

The Japanese seismic network record provides the EQ occurrences in stochastic time series, which completely masks subtle scale-dependent EQ phenomena originating from the D-B transition region [7]. PWs were used to describe the scale dependence as an equation of how principal stress changes generate significant earthquakes and megathrust earthquakes in every mesh size of about 4° by 5° in longitude and latitude throughout Japan [6, 7, 12–15]. The equation in each mesh describes consecutive EQ events as a virtual EQ particle motion of the unit mass experiencing crustal stress changes. The equation of EQ particle motion can identify the deterministic factors that control the significant and megathrust EQ genesis processes. The mesh size and shape can vary, and the mesh can be a small or large region covered by a seismic network [6].

Real-time monitoring of the genesis processes of megathrust and significant EQs observed in the GPS and seismic records can provide the deterministic predictions of the EQs and tsunamis as well as the hazard mitigation strategies. The PWs make the deterministic real-time monitoring of the overwhelmingly stochastic EQ phenomena possible. Using a deterministic model instead of a solely probabilistic approach is imperative in effective hazard mitigation. An improved understanding of EQ genesis processes and the deterministic factors controlling of significant and megathrust EQs could significantly enhance EQ and tsunami hazard mitigation efforts.

**2 A virtual EQ particle's stochastic motion**

Each EQ event in a mesh detected by the seismic network has the property of the focus (in latitude *LAT*, longitude *LON*, and depth *DEP*), its origin time (event time), and magnitude *MAG*. They are the so-called EQ source parameters. The interval between consecutive event times is the inter-event interval (*INT*), reflecting the stress state [16]. In the *c*-coordinate space (*c* = *LAT*, *LON*, *DEP*, *INT*, and *MAG*), an event is a virtual EQ particle of unit mass that emerges and moves to a new location at the next event. The movement discontinuously changes direction and speed like a Brownian particle [17]. Each *c*-component is a zigzagged non-time differentiable pathway [6, 7, 12–15];

$$\{c\} = \{d(c,1), d(c,2), \cdots, d(c,m), \cdots, d(c,N)\}. \qquad (1)$$

The $d(c, m)$ is the *c*-coordinate of the EQ particle's position at time *m*. The time is the chronological event index *m* ($m \geq$ 1 because no *INT* at *m* = 0). The $d(c, N)$ is the last observed EQ position. Denote a reference position, or a mean of $d(c, m)$ averaged over *N* by $<d(c, m)>_N$, and the relative change from each reference, $d(c, m) − <d(c, m)>_N$, by $d(c, m)$. Thus, $\{c\}$ may be a displacement time series from each reference. Index time *m* is not a stochastic quantity. A selection of *MAG* $\geq Mc$ ($Mc \approx 3.5$) for every mesh shows that $d(c, m)$ is a function of the principal stress components having a reduced stochastic noise level of about 15 ~ 25 % [7].

After every main event, each noisy displacement $d(c, m)$ shows an exponential growth rate, named a Lyapunov exponent [7]. The most prominent of all exponents, the largest exponent, is statistically distinct from surrogates created by randomly shuffling the chronological index $m$ in $\{c\}$, suggesting the significant event is deterministic chaos [7, 14]. The deterministic evidence buried in $\{c\}$ has a lasting memory of the significant event revealed with the Hurst exponents [7]. The evidence suggests some physical models exist on significant and megathrust EQ generations. However, the models found in seismic catalogs are all statistical, like the Omori law on the cumulative aftershocks [18], epidemic-type aftershock sequences (ETAS) [19, 20], and their asymmetry behavior [21].

The magnitude selection with $MAG \geq Mc$ ($Mc \approx 3.5$) shows position $d(c, m)$ having depth-dependent frequency-size distributions for $M = d(MAG, m)$ and $t = d(INT, m)$ [7]. Physical Wavelets [6–8] expect to observe similar scale-dependent seismogenic structures and processes of a radius of 120 km [22–25]. No other tools or operators can locate and extract the scale-dependent and deterministic evolution of the principal stress components to significant and megathrust EQs.

## 3 Physical Wavelets

Physical wavelets are powerful observational and mathematical operators for defining position (displacement), velocity, and acceleration on stochastic time series data $\{c\}$ to study the underlying dynamics [6–8]. The operators are the displacement-defining operator $DDW(t - \tau)$, velocity-defining operator $VDW(t - \tau)$, and acceleration-defining operator $ADW(t - \tau)$ at time $\tau$. They satisfy the time reversal property for displacement and their derivatives, enabling the definition of respective fundamental physical quantities. By taking the cross-correlation (or inner product in the case of vector representations) of each operator with the time series data $\{c\}$ that is non-differentiable in time, we can define displacement, velocity, and acceleration.

The displacement $D(c, \tau)$ is defined using the cross-correlation of $DDW(t - \tau)$ with $\{c\}$:

$$D(c,\tau) = \int_{-\infty}^{+\infty} \{c\} DDW(t-\tau) dt = [1/(2w+1)] \sum_{m=-w}^{w} d(c, \tau + m). \tag{2}$$

Equation (2) shows that $D(c, \tau)$ is defined by the moving average of $2w + 1$ values of the stochastic particle displacement $d(c, m)$. Velocity $V(c, \tau)$ and acceleration $A(c, \tau)$ are defined using $VDW(t - \tau)$ and $ADW(t - \tau)$, respectively, resulting in replacing the time differentiation of $D(c, \tau)$ by taking the difference at $s$ events separation:

$$V(c,\tau) = \int_{-\infty}^{+\infty} \{c\} VDW(t-\tau) dt = [D(c, \tau + s/2) - D(c, \tau - s/2)]/s \tag{3}$$

and

$$A(c,\tau) = \int_{-\infty}^{+\infty} \{c\} ADW(t-\tau) dt = [V(c, \tau + s/2) - V(c, \tau - s/2)]/s$$
$$= [D(c, \tau + s) - 2D(c, \tau) + D(c, \tau - s)]/s^2. \tag{4}$$

The time-reversal operation changes $\tau$ to $-\tau$ and confirms that $D(c, -\tau) = D(c, \tau)$, $V(c, -\tau) = -V(c, \tau)$, and $A(c, -\tau) = A(c, \tau)$. Thus, they exhibit the time-differential properties for the differentiable $D(c, \tau)$ and $V(c, \tau)$ obtained from the stochastic $d(c, j)$ of $\{c\}$. The relationships between $D(c, \tau)$, $V(c, \tau)$, and $A(c, \tau)$ are the equations of averaged EQ particle's stochastic motion, which may carry periodically fluctuating components and trends embedded in $\{c\}$ [6–8].

The extraction of specific fluctuations embedded in $\{c\}$ is significant if the mutual correlation between Physical Wavelets and $\{c\}$ is strong. The extracted physical quantities are Eqs. (2), (3), and (4), representing a low pass filtered, a bandpass filtered, and another bandpass filtered quantity, respectively. The parameters $w$ and $s$ can be any integer with which to filter out the selected frequency components of $D(c, \tau)$, $V(c, \tau)$, and $A(c, \tau)$. The cut-off frequencies in the fluctuation (frequency $f$ in 1/events) domain are respectively $(4s)^{-1}$ and $(4s/3)^{-1}$ for high-pass and low-pass filters [6–8]. Thus, $A(c, \tau)$ oscillates with a periodic frequency component of $f \approx (2s)^{-1}$, extracted from time series $\{c\}$. We note that time $\tau$ lags from the current time $j$ because $\tau = j - w - s$. Thus, real-time observation of the EQ average motion is at time $j = \tau + w + s$.

# 4 Significant EQ genesis processes and predictions

Real-time monitoring of the significant EQ genesis processes observed in the seismic and GPS records can provide the deterministic EQ predictions and hazard mitigation strategies.

## 4.1 Oscillatory CQK and CQT observed in seismic records

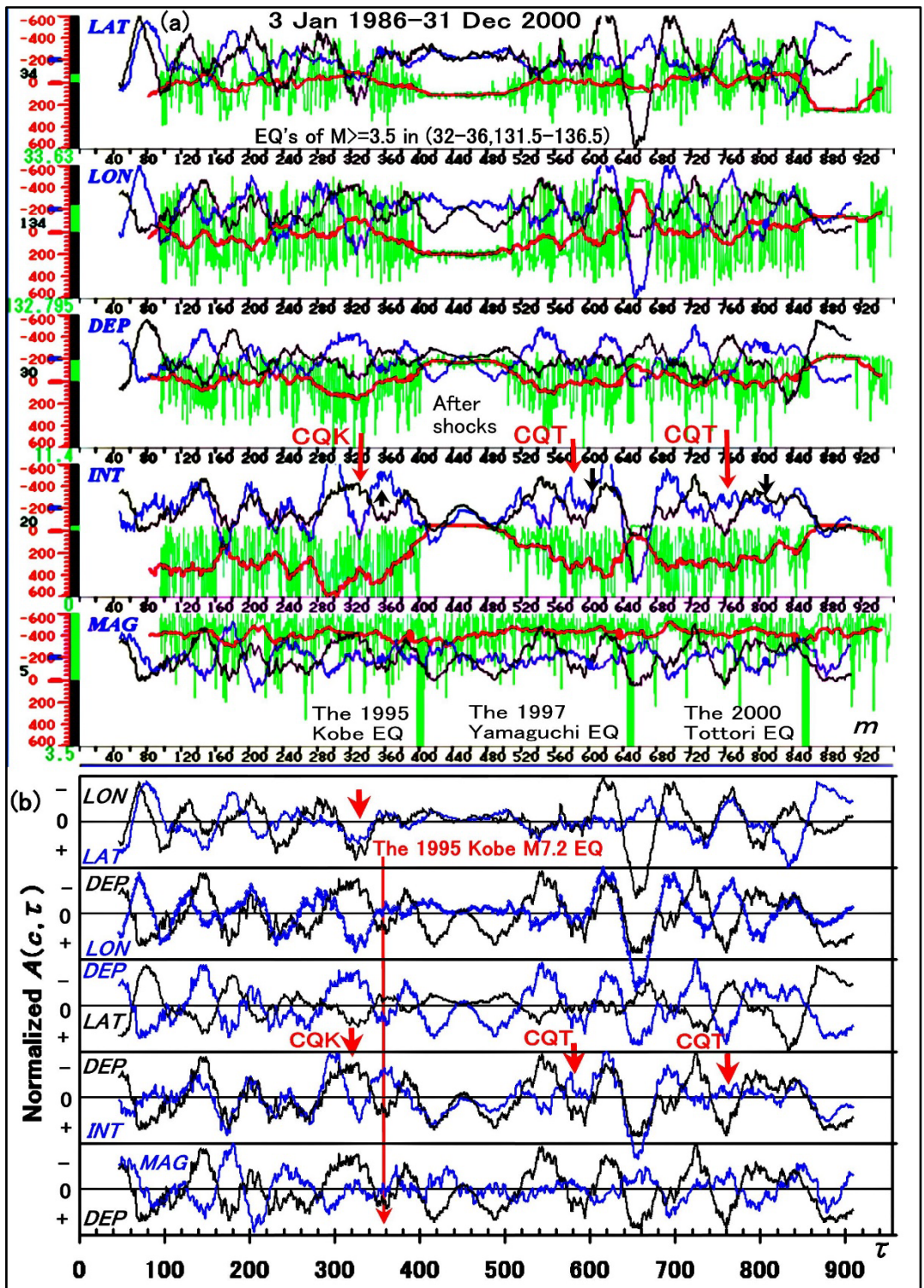

**Figure 1. CQK and CQT**

Figure 1 is modified and reproduced from references [6, 7]. (a) An EQ particle motion with $MAG \geq 3.5$ and $DEP \leq 300$ km in a small mesh region of $LAT = 32°–36°$ N and $LON = 131.5°–136.5°$ E. The shallow particle motion in the EQ $c$ coordinate space whose $c$ axis is the vertical axis of $LAT$, $LON$, $DEP$, $INT$, and $MAG$ on the left. All the horizontal scales are in chronological event index $m$, which shares other time $t$ and $\tau$ where $m = t + w$ and $t = \tau + s$ ($w = 12$ and $s = 35$). The date at $m = 0$ and 1 is 3 Jan 1986 and 5 Jan 1986, respectively. The $d(c, m)$ in green is the relative position from each graphical reference of 34° N, 133° E, 30 km, 20 hours, and 5. They are all at zeroes of the manometer-like scales on the left. The scale magnification is 200 times for $LAT$ and $LON$, 10 times for *DEP*, 2 times for *INT*, and 400 times for *MAG*. For example, the *LAT*, *LON*, and *DEP* at scale − 200 correspond to 33° N, 132° E, and 10 km, respectively. The *MAG* range is from 3.5 (− 600) to 6.5 (600), so it saturates above 6.5. Each manometer column illustrates a variation of $d(c, m)$ with its digital reading during monitoring. The last readings at $m = 956$ (31 Dec 2000) are *LAT* = 33.63° N, *LON* = 132.795° E, *DEP* = 11.4 km, *INT* = 0 (0.2165) hours, and *MAG* = 3.5. The positive direction is downward from each reference point. The relative position $D(c, t)$ and acceleration $A(c, \tau)$ exhibit the periodic fluctuations of about 70 (2s) events. The $D(c, t)$ is red, and $A(c, \tau)$ is blue and black. From the top axis, the black $A(c, \tau)$ is $c =$ *LON, DEP, LAT, DEP*, and *DEP*. Their relative amplitudes are from each origin of the blue bar marked at −200 on the left scale. The $d(c, m)$, $D(c, t)$, and $A(c, \tau)$ become bold at the events of *MAG* ≥ 6. The bold lines of $d(MAG, m)$ with black arrows have the EQ names of the 1995 Kobe M7.2 (at $m =$ 402, 17 Jan 1995), the 1997 Yamaguchi M6.6 (at $m = 649$, 25 Jun 1997), and the 2000 Tottori M7.2 (at $m = 856$, 6 Oct 2000). The black arrows on $A(INT, \tau)$ point to their temporal locations. (b) Normalized pair of $A(c, \tau)$. Each pair of blue and black $A(c, \tau)$ is normalized to its maximum $A(c, \tau)$ in the − region. Thus, every amplitude of $A(c, \tau) < 0$ is within − one. The first and last 47 events for $A(c, \tau)$ are not obtained because of $m = \tau + s + w$ ($w = 12$ and $s = 35$). The 1995 Kobe M7.2 EQ is at the long-downward arrow. The short arrows in *DEP–INT* row are at CQK and CQT. Blue $A(LAT, \tau)$ and black $A(LON, \tau)$, pointed to by another short arrow above CQK, show $A(LON, \tau) > A(LAT, \tau) > 0$.

With a selection of *MAG* ≥ *Mc* (*Mc* ≈ 3.5) in a small mesh of about 4° by 5° in longitude and latitude, the EQ particle motion, averaged over $2w + 1$ events ($w \approx 7 \sim 17$), is under a restoring force of $F(c, \tau) = A(c, \tau) \approx - K(c) \times D(c, \tau)$ [6, 7]. Each $A(c, \tau)$ has a periodicity of about 2s. The positive constant $K(c)$ is a weak function of time $\tau$. The $A(c, \tau)$ and $D(c, \tau)$ are the functions of noise-free three principal stress components. As the EQ average movement approaches an imminent significant EQ, $A(c, \tau)$ among $c =$ *DEP*, *INT*, and *MAG* establishes two phase-inversions between oscillatory $A(DEP, \tau)$ and $A(INT, \tau)$ with the negative amplitude of $A(MAG, \tau)$ [6, 7, 12–15]. An inversion with positive $A(INT, \tau)$ and negative $A(DEP, \tau)$ is CQK, named after the 1995 Kobe M 7.2, as in Fig. 1. Another inversion with negative $A(INT, \tau)$ and positive $A(DEP, \tau)$ is CQT after the 2000 Tottori M 7.2. The CQ stands for Critical Quiescence, K for Kobe, and T for Tottori.

The CQK and CQT in terms of $D(c, \tau)$ are CQKD and CQTD, as in Fig. 2. The significant EQ having CQK or CQT ruptures in the periodic motion with the expected EQ source parameters; focus, fault movement, size, and rupturing event time $m$ [6, 7, 12–15]. The real-time monitoring of strain-energy cycles in section 7 converts the event index time to the rupturing date and time [6, 7, 12–15]. Every significant EQ and swarm (M > about 6) exhibits either CQK or CQT throughout meshes in Japan, including the 2011 Tohoku M9 with CQK [6]. A few exceptions exist to the significant EQs with the preceding medium-size swarms (M ≈ 5) that masked CQK or CQT [6]. However, resizing the mesh reveals the CQK or CQT and even detects the medium-size swarms as an isolated CQT process. The EQ swarms in a mesh are all CQTs [6]. The CQK and CQT have seismogenic structures with upward (to the Earth's surface) and downward stress loading to the hanging wall at fault surfaces, respectively [6, 7]. The CQKs are all in the low Coda Q spots, whereas the CQTs are in the high Coda Q spots throughout the Coda Q map of Japan [6, 26].

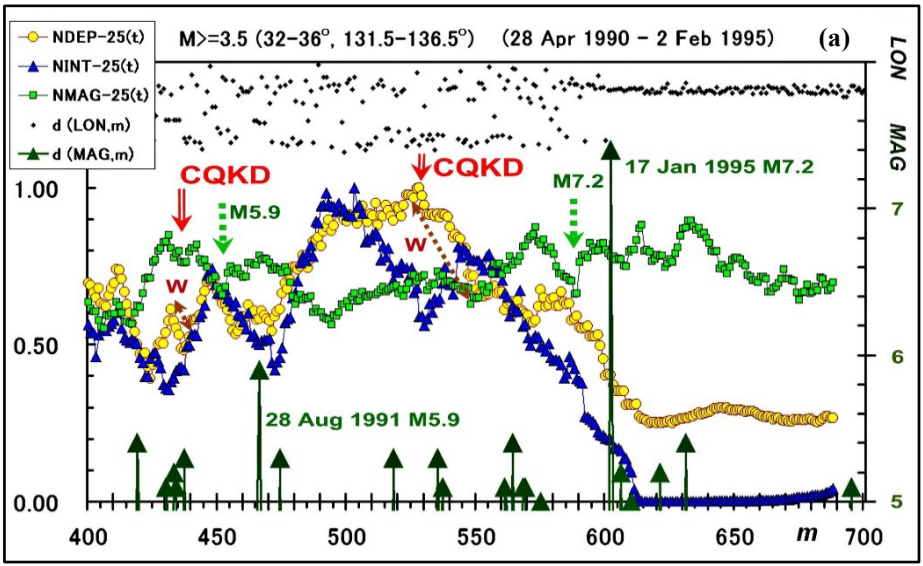

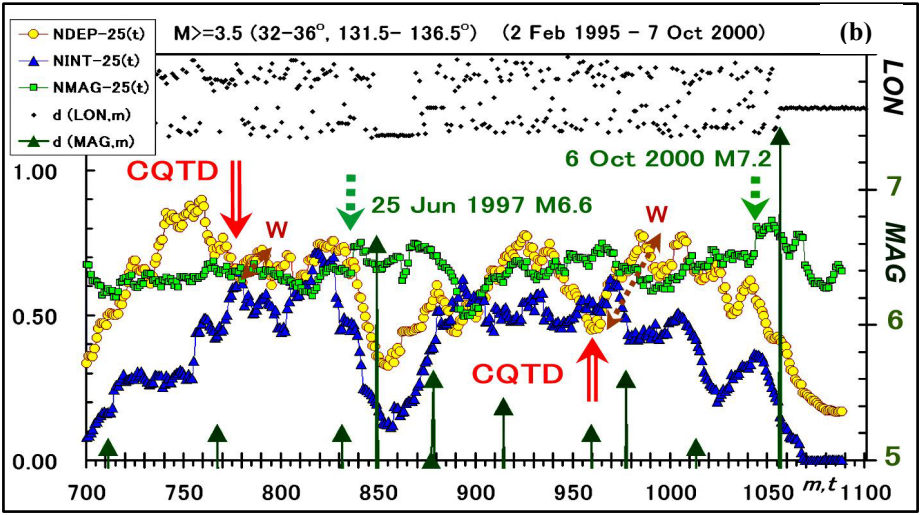

**Figure 2. CQKD and CQTD**

Figure 2 is modified and reproduced from references [6, 7]. (a) The CQKD of $D(DEP, t)$, $D(INT, t)$, and $[D(MAG, t) − 3]$ with $2w + 1 = 25$ events in Eq. (2) are normalized to NDEP–25(t), NINT–25(t) and NMAG–25(t) with the past maximum values of 47.16 km, 331.33 h, and 1.4, respectively. At the Kobe CQKD of $t = 528$ (28 Jun 1993), $D(DEP, t)$, $D(INT, t)$, and $D(MAG, t)$ are 47.16 km, 195.48 h, and 4.0, respectively. The MAG scales of $d(MAG, m)$ are on the right, for which $MAG \geq 5$ is in up-arrows. The $d(LON, m)$ is relative on the righthand scale. Index $m = 1$ starts on 8 Jan 1983, not 5 Jan 1986, as in Fig. 1a. The 1995 Kobe event in dot-arrow is at $t = 590$ ($m = 602$). The fault width $W$ is on a depth segment of NDEP-25(t). The estimated $W = 20$ km is in section 4.3. Figure part (a) shows that the M5.9 with $w = 5.9$ km is at $t = 454$ ($m = 466$, 26 Aug 1991). The $D(DEP, t)$ oscillation at the CQKD has about 14-event periodicity that the $A(DEP, \tau)$ of 70-event periodicity filtered out, resulting in $A(DEP, \tau)$ and $A(INT, \tau)$ in phase. Thus, we do not observe the M5.9 CQK [6]. (b) The CQTD for the 1997 Yamaguchi M6.6 (moment magnitude $M_W = 5.9$, $W = 5.5$ km) is at $t = 837$ ($m = 849$, 25 Jun 1997), and the 2000 Tottori M7.2 ($M_W = 6.8$, $W = 15$ km) is at $t = 1044$ ($m = 1056$, 6 Oct 2000).

### 4.2 Particular time $\tau a$, $\tau b$, and $\tau r$

The time $\tau a$, $\tau b$, and $\tau r$ are for $c = INT$ and $DEP$ during CQK and CQT [6, 7]. Time $\tau a$ is at the first positive (or negative) $A(c, \tau)$ peak amplitude. The $\tau b$ is when $A(c, \tau)$ is zero after the first peak. The $\tau r$ is at the second $A(c, \tau)$ peak. The imminent significant event ruptures at time $\tau r$ for $c = INT$ or $DEP$, as in Fig. 1 [6, 7, 12–15]. Unlike perfect inversions of CQK (the 1995 Kobe M7.2) and CQT (the 2004 Chuetsu M6.8 [6]), there are partial phase inversions between $A(INT, \tau)$ and $A(DEP, \tau)$. For such cases, some $\tau r$ has a delay to the third $A(c, \tau)$ peak [6]. To predict the expected peak in $A(c, \tau)$ at current time $j$ where time $\tau = j − w − s$, we use the periodicity in $A(c, \tau)$, or we locate a sharp change in $D(c, \tau)$ that will expect a peak in $A(c, \tau)$ because of time $\tau = j − w$ for $D(c, \tau)$ [6].

### 4.3 Equations for fault movement, size, and magnitude

We assume a significant EQ has a fault plane of width $W$ and length $L$ in km. The restoring force $F(c, \tau)$ with $c = LAT, LON$, and $DEP$ generates the shear stress exceeding the critical failure value of the fault plane during CQK or CQT [6, 7]. The horizontal force components to the Earth's surface are $F(LAT, \tau a)$ and $F(LON, \tau a)$. The expected fault length $L$ has the components $D(LAT, \tau a)$ and $D(LON, \tau a)$ in degrees [7]. The vertical force component with $F(DEP, \tau a) < 0$ is upward (to the Earth's surface) for CQK. The $F(DEP, \tau a) > 0$ is downward (from the Earth's surface) for CQT. Figure 2 shows an example of expected fault width $W \approx |D(DEP, \tau a)|$ [6, 7].

As for the 1995 Kobe M7.2 CQK, the horizontal force points clockwise 60° from N (north) because $F(LAT, \tau a)$ and $F(LON, \tau a)$ are proportional to $(LAT, LON) = (0.25°, 0.53°)$ [7]. The $F(DEP, \tau a)$ points upward. An arrow in Fig. 1b also shows a magnitude relation of $A(LON, \tau) > A(LAT, \tau) > 0$ during the entire CQK, suggesting the net force direction of about 68° (= 45° + 45°/2) clockwise from N. Thus, $F(c, \tau a)$ will induce the shear stress to move the hanging wall toward the net force direction in a right-lateral strike-slip with an upward dip-slip component. The Kobe EQ focal mechanism had (Strike, Dip, Rake) = (233°, 86°, 167°) [27, 28]. The 167° rake indicates that the shear stress moved the hanging wall opposite the strike at 53° (= 233° − 180°) clockwise from N, having 66° (= 233°−167°) clockwise from N and upward by 13° relative to the reference strike in the fault plane. The observation supports the expected fault movement by $F(c, \tau a)$. Furthermore, the displaced magnitudes are $|D(DEP, \tau a)| = 20$ km, $|D(LAT, \tau a)| = |− 0.25°|$ and $|D(LON, \tau a)| =$

|−0.53°|, having the total length of 56.4 km in agreement with the aftershock distribution of $W$ = 20 km and $L$ = 77 km [29].

The 1997 Yamaguchi M6.6 (at $m$ = 649, 25 Jun 1997, in Fig. 1) had CQT having the focal mechanism of (Strike, Dip, Rake) = (55°, 83°, −153°) [28]. The hanging wall slipped opposite the reference strike at 235 ° clockwise from N, having 208° (= 55° + 153°) clockwise from N with a downward dip-slip of 27° relative to the reference in the fault plane. The $A(LON, \tau a) \approx A(LAT, \tau a) < 0$ at $\tau a$ from $A(DEP, \tau a)$ in Fig. 1b suggests the net force direction of 225° (= 180° + 45°) clockwise from N to move the hanging wall. Figure 2b shows $W \approx |D(DEP, \tau a)|$ = 5.5 km. The aftershock distribution was $(W, L)$ = (10 km, 10km) [30].

Similarly, the 2000 Tottori M7.2 (at $m$ = 856, 6 Oct 2000, in Fig. 1) was CQT with (Strike, Dip, Rake) = (145°, 71°, −11°) [28]. The focal mechanism shows the hanging wall slipped toward the reference strike of 145 ° clockwise from N, having 156° clockwise from N with a downward dip-slip of 11° from the reference in the fault plane. The $A(LON, \tau a) \approx A(LAT, \tau a) < 0$ at $\tau a$ of $A(INT, \tau a)$ suggests the net force pointing clockwise 225° from N moved the hanging wall, different from the slip direction. However, a sharp transition from $A(LON, \tau a) \approx A(LAT, \tau a) < 0$ to $A(LON, \tau) > 0$ and $A(LAT, \tau) \approx 0$ for $\tau a < \tau < \tau r$ suggests the net force clockwise direction from N is 90° ∼ 225°, which covers the slip direction of 145°. The fault width $W$ in Fig. 2b shows $|D(DEP, \tau a)|$ = 15 km in agreement with the aftershock distribution of $(W, L)$ = (15km, 30km) [31].

The observed length $L$ and width $W$ in km during CQK and CQT become an imminent EQ's planar-fault size. The significant EQ's magnitude can be estimated by an empirical $M = \log S + 3.9$ ($S = L \times W$) [18]. As for the Kobe CQK observation, the predicted $M$ is 6.9. The seismological observation of $M$ is 6.9 and 7.2 for the moment and the JMA magnitude [32], respectively. The estimation of JMA's $M$ with an assumption of $L = 2W$ ($W \approx |D(DEP, \tau a)|$) agrees with the observed $M$ for most significant events throughout Japan [6].

### 4.4 Equations for rupture time

The $D(INT, \tau r)$ and $A(INT, \tau r)$ show an expected rupture-time $\tau r$ in $m = \tau r + s + w$ within one or two event time accuracy [6, 7, 12–15]. For example, the Kobe CQK showed $\tau r$ = 19 events on 24 Oct 1994 [6, 7]. The Kobe EQ event ruptured in 19 events on 17 Jan 1995. The $D(DEP, \tau r)$ and $A(DEP, \tau r)$ may replace the corresponding $D(INT, \tau r)$ and $A(INT, \tau r)$ to predict the rupture time. The conversion from the event time $\tau r$ to the date and time requires an average rate of events obtained by a strain energy cycle in section 7 [6, 7].

Some CQK events, having a significant upward dip-slip component and $MAG$ < about 6, have considerable time delays from the expected rupture-time $\tau r$. The delay is due to the fault failure's insufficient upward share stress loading. Thus, another upward loading at a new $\tau r$ of oscillatory CQK is necessary [6]. Other CQK events have $MAG \approx 6$ during the CQK process, and then CQT follows. The 2003 Off-Tokachi megathrust M8 EQ was such a case, and the M8 EQ ruptured at time $\tau r$ of CQT [14].

The final rupture time prediction becomes within a day accuracy by real-time monitoring of the strain energy accumulation and release cycles, as in section 7.

### 4.5 Equations for the focus

As for the equation of focus during CQK and CQT, the $A(c, t)$ with $c$ = LAT, LON, and DEP may make the corresponding $D(c, t)$ have the most straightforward linear extrapolations to the expected rupture-time [6, 7, 12–15]. For example, the linear prediction shows (34.53°, 135.18°, 20.8 km) in (LAT, LON, DEP) for the Kobe EQ, which matches the (34.595°, 135.038°, 16.06 km) given in the JMA catalog [4].

### 4.6 Real-time monitoring of CQK and CQT

The CQK and CQT are unique amplitude and phase relationships among $A(DEP, \tau)$, $A(INT, \tau)$, and $A(MAG, \tau)$ at time $\tau$, for which a time shift to the current time $j$ (= $\tau + w + s$) does not change the property of the CQK and CQT. Thus, the real-time and automatic detection of CQK and CQT is also available by monitoring the EQ particle's movement power. The power is the time-rate change of the kinetic energy defined by $PW(c, \tau) = V(c, \tau) \times A(c, \tau)$. For the real-time monitoring at the current time $j = \tau + w + s$ in $A(c, \tau)$, the power is $PW(c, j) = V(c, j) \times A(c, j)$, as in section 6 [33].

### 4.7 $PW(c, j)$ monitoring of anomalous lunar tidal loadings in real-time

Suppose an anomaly in the crustal responses to an external force is a change comparable to or less than the amplitude fluctuations in $\{c\}$. In that case, it will be indistinguishable from the background trends and fluctuations. It may appear as subtle changes in their phases and magnitudes. An automated and real-time power monitoring, $PW(c, j)$, is a well-established and patented method to locate anomalies and prevent real-time industrial system failures and disasters [33]. Some other applications are to detect and locate such anomalies as the most critical information for bio-medical engineering [34, 35], industrial [36], and physical [37, 38] systems.

The daily displacements at a GPS station are $\{c\}$ of Eq. (1) in a Cartesian $c$-coordinate system in right-handed $(E, N, h)$, where $c$ = E (eastward), N (northward), and h (upward). Each displacement is noisy and non-time-differentiable;

however, {c} contains the crustal responses to the lunar tidal force loadings. Extracting a tidal force loading, for example, fortnightly (bi-weekly) loading from daily displacement series {c = h} under the ± 20 mm noise requires $w \approx 2$ and $s \approx 7$ to obtain $D(c, \tau)$, $V(c, \tau)$, and $A(c, \tau)$. The periodic responses to the lunar loadings expect standard unless the regional frictional failure stress state is in an imminent CQK or CQT. The $PW(h, j)$ detects unusual crustal responses to the fortnightly tidal loading about two weeks before EQs with $MAG \geq$ about 5 [39]. Thus, the GPS $PW(c, j)$ monitoring of the periodic lunar tidal force loadings is a real-time supplementary tool to follow CQK or CQT.

As for the synodic loading, the $PW(c, j)$ detection requires $w \approx 7$ and $s \approx 20$ to obtain $D(c, \tau)$, $V(c, \tau)$, and $A(c, \tau)$ [8]. Appendix B in the reference [8] shows the detailed synodic monitoring of many significant EQ events of M6.9, M6.4, M7.9, M8.1, M7.3, and M6.9. The M8.1 was west off-Ogasawara, 681.7 km deep, one of Wadati-Benioff zone EQs.

## 5 Megathrust EQ genesis processes and predictions

Real-time monitoring of the genesis processes of megathrust EQs observed in the seismic and GPS records can provide deterministic predictions of EQs and tsunamis and enable hazard mitigation strategies up to the event occurrences.

### 5.1 CQK and CQT processes observed in seismic records

The 2011 Tohoku M9 EQ genesis processes in a Tohoku mesh (36° – 40°, 138° – 143°) had the CQK and CQKD established by 24 Feb 2011, 16 days before the M9 event [6]. There are neighboring meshes [6, 7, 14]; mesh (32.5° – 38°, 136.5° – 142°) having CQT and CQTD, and mesh (36° – 40°, 136° – 140°) having CQT and CQTD on the same date [6]. The three significant EQ genesis processes suggest that they are all coupled with the M9 event on 11 Mar [6]. The Tohoku CQK shows $A(LON, \tau a) > 0$ and $A(LAT, \tau a) < 0$ with $| A(LON, \tau a) | > | A(LAT, \tau a) |$, suggesting that the horizontal force will induce share stress making the hanging wall slip to 113° (= 90° + about 45° / 2) clockwise from N. The slip direction is perpendicular to the east coastline in Fig. 3. The force $F(c, \tau)$ with $c = LAT$, $LON$, and $DEP$ will make the hanging wall (the Tohoku east coast) slip upward. It will generate a reverse fault as in Fig. 4c. The Tohoku M9 EQ had the focal mechanism of (Strike, Dip, Rake) = (193°, 10°, 79°) [28]. The observed slipping direction of the hanging wall is 114° (= 193° − 79°) clockwise from N in the fault plane having a shallow dip angle of 10°. The rake of the dip-slip vector was almost perpendicular to the reference strike along the east coastline. Thus, the focal mechanism agreed with the predicted fault movement. The CQK seismogenic structure is also consistent with the east coastline of the low Coda Q map [6, 7, 26]. As for the prediction of the focus and rupturing date made on 24 Feb 2011, they were (38.24°, 142.82°, 19 km) in (LAT, LON, DEP), and in 3 events (9 Mar 2011) by $A(DEP, tr)$ or 6 events (21 Mar 2011) by $A(INT, tr)$ [6] in agreement with the observed focus of (38.10°, 142.85°, 24 km) on 11 Mar 2011. However, the Tohoku CQK and CQKD showed that the predicted fault width $W = 17$ km was much less than $W = 200$ km in Fig. 3 and Fig. A1 [6–8]. The mesh size for significant EQ predictions appears inappropriate for the megathrust EQ's fault size prediction.

Another megathrust EQ was the 2003 Off-Tokachi M8 EQ in Fig. 3. The M8 EQ source parameters are focus = (41.78°, 144.08°, 42 km), origin time = 04:50 on 27 Sep 2003, and focal mechanism = (260°, 6°, 144°) in (Strike, Dip, Rake) [28]. The Tokachi M8 had a consecutive CQK and CQT in the Hokkaido mesh (40.5° – 45.5°, 141.5° – 145.5°) [14]. The EQ prediction with CQT made on 31 Aug 2003 was the fault width $W = 25$ km ($M = 7.0$ with the assumption of $L = 2W$), the focus of (42.38°, 143.73°, 66.57 km), and the rupturing date and time of 27 Sep 2003 06:00 (8 events = 28.4 days from 31 Aug 2003) [14]. The Tokachi CQT shows $A(LON, \tau a) > 0$, $A(LAT, \tau a) < 0$, and $| A(LON, \tau a) | > | A(LAT, \tau a) |$, suggesting that the horizontal force made the hanging wall (Tokachi east coast) slip to 113° (= 90° + about 45° / 2) clockwise from N in agreement with the observed rake of dip-slip vector, 116° (= 260° − 144°) clockwise from N in the fault plane at a shallow dip angle of 6°. However, the vertical force with $A(DEP, \tau a) > 0$ points downward from the Earth's surface, contradicting the rake of 144°. The dip-slip vector points upward 36° from the reversed reference strike (from the Tokachi east coast from SW to NE, as in Fig. 3). Although no Coda Q map is available for the Hokkaido area [26], the Tokachi M8 had an exceptional CQT with preceding CQK having $A(MAG, ta) \approx 0$. The Tokachi M8 generated the highest tsunami of 2.25 m [40], much less than the Tohoku M9 tsunami above 9.3 m [40]. The Tohoku CQK had an upward dip-slip vector directed to 79° relative to the reference strike. The reference is the Tohoku east coastline from NNE to SSW, as in Fig. 3. Thus, the megathrust EQ with a consecutive CQK (weak) and CQT may account for a small angle of 36°-dip-slip vector to generate the 2.25 m height tsunami. The aftershock distribution shows the fault size of $(W, L) = (160$ km, $140$ km$)$ [41].

### 5.2 GPS observations

The {c} of Eq. (1) represents the GPS displacements in a Cartesian $c$ coordinate system in right-handed ($E, N, h$), where $c = E$ (west to east), $N$ (south to north), and $h$ (down to up). Each displacement is noisy and non-differentiable in time; however, Physical Wavelets can define smooth $D(c, \tau)$, $V(c, \tau)$, $A(c, \tau)$, $D(h, \tau) - V(h, \tau)$ path, and $D(h, \tau) - A(h, \tau)$ path to quantify the underlying crustal dynamics observed as {c}. The GPS observation shows that the 2011 Tohoku M9 EQ and tsunami genesis processes resulted from complex interactions between the subducting and overriding plates, causing bulge-bending deformation of the Tohoku crust [6, 8]. This bulge deformation is a transition from the regular deformation expected for the past three hundred years [8]. The fifteen-month genesis process of the megathrust EQ and the last three-month tsunami genesis process can provide the real-time prediction of the EQ and tsunami and the hazard mitigation strategies up to the event occurrences [6, 8].

**5.2.1 Tohoku subduction zone**

The foreshocks, aftershocks, and vertical co-seismic shifts of the Tohoku M9 EQ are shown in Fig. 3.

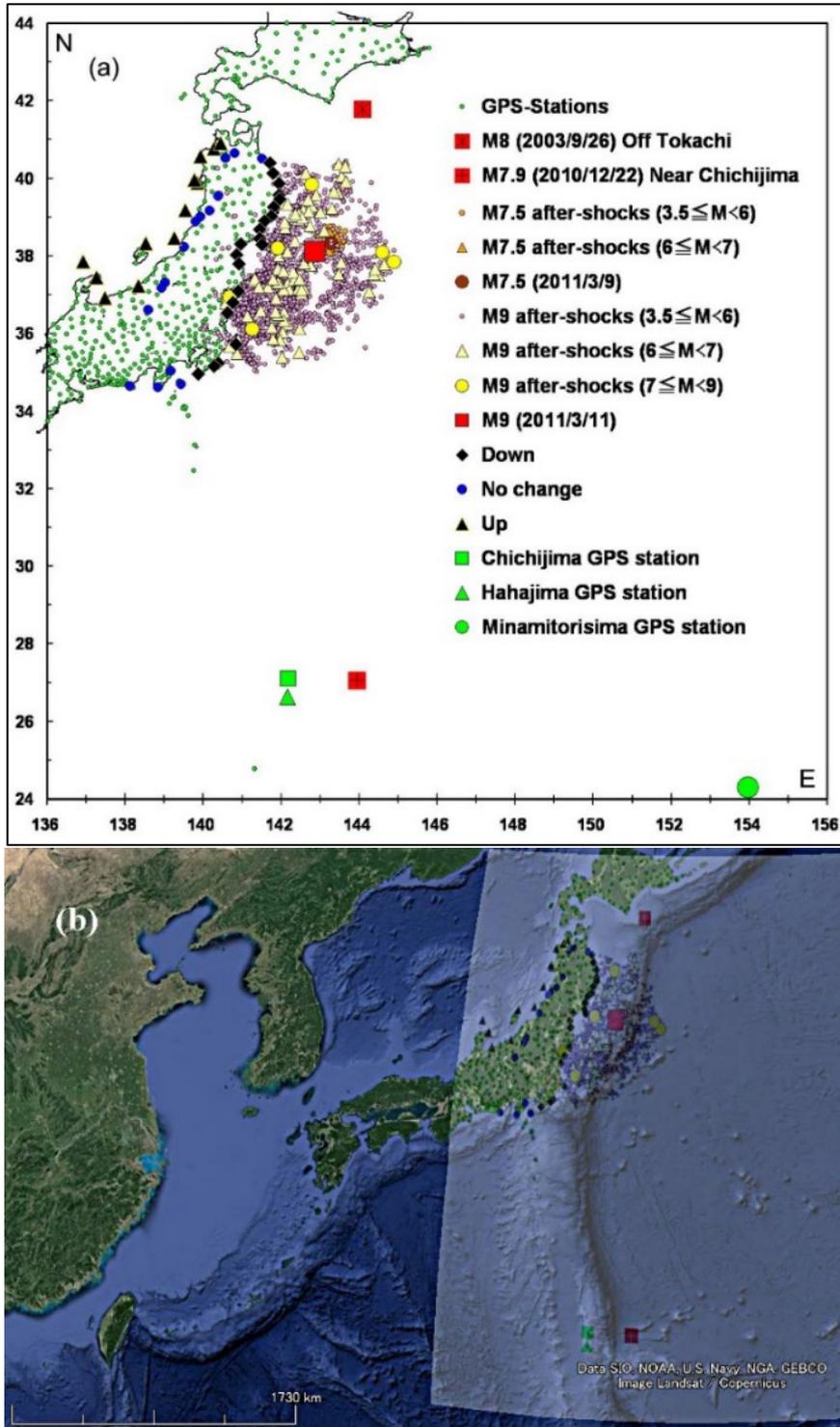

**Figure 3. GPS stations, vertical co-seismic displacement, foreshocks, and aftershocks of the M9 EQ**
Figure 3 is modified and reproduced from references [6–8]. Magnified graphs and a map are in Appendix. (a) The EQ source parameters are from the JMA's unified focus catalogs [4]. The EQ magnitude M is JMA's magnitude. The off Tokachi M8 (2003/9/23) EQ [12] was the megathrust EQ in section 5.1. The M7.5 (2011/3/9) EQ is a foreshock of the Tohoku M9 EQ (2011/3/11).

The M9 had the focus of (38.1006°N, 142.8517°E, 24 km), and the reverse faulting of (Strike, Dip, Rake) = (193°, 10°, 79°) [28]. The vertical co-seismic displacements over the 500 km distance are downward (Down), upward (Up), and no change (No change) displacement at each GPS station. Three GPS stations above 38 degrees north latitude-line, used for Fig. 4, are Ryoutsu2 (Up) at (38.0633° N, 138.4717° E, west coast), Murakami (No change) at (38.2307° N, 139.5069° E), and Onagawa (Down) at (38.4492° N, 141.4412° E, east coast). Their observations show the Tohoku bulge-bending deformation processes as in Figs. 4 and 5. The M9 EQ (2011/3/11) is on the 38-degree N line. The GPS stations in the Northwest Pacific Ocean are Chichijima, Hahajima, and Minamitorishima. Chichijima and Hahajima stations are on the subducting western edge of the northwestern Pacific Plate, whereas Minamitorishima station is on the northwestern Pacific Plate (far right below). The Chichijima, Hahajima, and Minamitorishima stations observed the identical abnormal oceanic plate motion coupled with the bulge-bending deformation along the Tohoku east coast. As in Fig. 6, the Pacific Plate's abnormal westward motion triggered the M7.9 (2010/12/22) EQ of the normal faulting in the plate slab near Chichijima. The EQ had the focal mechanism of (Strike, Dip, Rake) = (109°, 46°, −131°) [28]. The hanging wall slipped to 240° clockwise from N, along the plate's westward motion. (b) A google earth map with Fig.3a overlaid.

The GPS stations [5] depicted in Fig. 3a and Fig. A1 captured the co-seismic downward and upward displacement along the east and west coasts, as well as no displacement along a ridge on Tohoku. The distribution of these observations suggests that the rectangular fault surface of the earthquake was about 500 km long and 200 km wide. The earthquake's occurrence decoupled the overriding Tohoku crust (the eastern edge of the overriding continental plate) from the subducting western edge of the northwestern Pacific Plate. Thus, the GPS observation suggests that the correct fault length estimation should be made by analyzing the daily displacement recorded at the GPS stations in Tohoku and the Northwest Pacific Ocean.

As for the Northwest Pacific Ocean, Fig. 3 shows three islands having GPS stations: Minamitorishima, Hahajima, and Chichijima. The Minamitorishima station is on the northwestern Pacific Plate. The Chichijima and Hahajima stations are located below the Japan Trench and on the western edge of the Ogasawara Plateau [5]. Their qualitatively identical GPS displacement time series $\{c\}$ suggests that Chichijima and Hahajima stations are on the western edge of the subducting northwestern Pacific Plate, and Minamitorishima station is under the northwestern Pacific Plate's motion. The GSI permanently closed Chichijima station (27.0956° N, 142.1846° E) on 9 Mar 2011. The last observation was on 8 Mar, three days before the Tohoku M9 EQ on 11 Mar 2011. Chichijima-A station (27.0675° N, 142.1950° E) started its observation on 4 Dec 2007, and replaced the Chichijima station.

### 5.2.2 A megathrust EQ and Tsunami genesis processes

The Tohoku M9 EQ and the resulting tsunami were not caused by the elastic rebound of the east coast subsiding with the subducting oceanic plate coupled with the fault [6, 8]. Instead, they resulted from complex interactions between the subducting and overriding plates, which caused bulge-bending deformation over the Tohoku subduction zone. The bulge-bending deformation refers to the gradual bending of the overriding plate of Tohoku by the eastward continental plate-driving force, causing the over-riding crust of Tohoku to bulge upwards and downwards by a few millimeters. This deformation evolved from the expected regular deformation that had been occurring for the past three hundred years [8].

### 5.2.2.1 Distinctive phases of the Tohoku crustal deformation

The GPS observation detected the onset of the bulge-bending deformation, which grew in three distinct phases: an initial phase of 6 months with gradual subsidence of 1 to 2.8 mm across the Tohoku region, followed by a transitional phase of 6 months with further gradual subsidence of 3.3 mm on the east coast, and a final phase of 3 months with an upheaval growth of 1 to 3 mm across Tohoku until the 2011 EQ event.

### 5.2.2.2 Regular deformation (Before January 2010)

Figure 4a illustrates the regular deformation of the dotted line over Tohoku with the subducting oceanic plate moving westward, causing the east coast to subside (represented by a westward arrow) and the overriding continental plate moving eastward, causing the west coast to move upward (represented by an eastward arrow). For instance, the Onagawa station (in Fig. 3) had a subsidence rate of 6 mm per year, while the Ryoutsu2 station (in Fig. 3) had an upward displacement rate of 1.5 mm per year. The subducting northwestern Pacific Plate moved westward at an average rate of 0.1 mm per day and an average northward movement of 0.03 mm per day. The rate of westward movement is approximately equivalent to the growth rate of a fingernail.

### 5.2.2.3 Initial phase (January 2010 to June 2010)

One month prior to January 2010, the deformation property of the west coast changed from non-elastic to elastic, leading to the generation of a restoring force on the compressed west coast. This property change of the west coast marks the onset of an underlying bulge-bending deformation that persisted for 15 months, resulting in gradual subsidence and upheaval growth of a few millimeters over a 500 km stretch along the east and west coasts.

In January 2010, the westward-restoring force of the west coast, compressed by the overriding continental plate-driving force, caused a gradual subsidence of 1 to 2.8 mm across the Tohoku region. By June 2010, the subsidence had reached 1 mm on the west coast, 1.5 mm on the top, and 2.8 mm on the east coast. The east coast subsidence was a pre-transition process from regular to bulge-bending deformation.

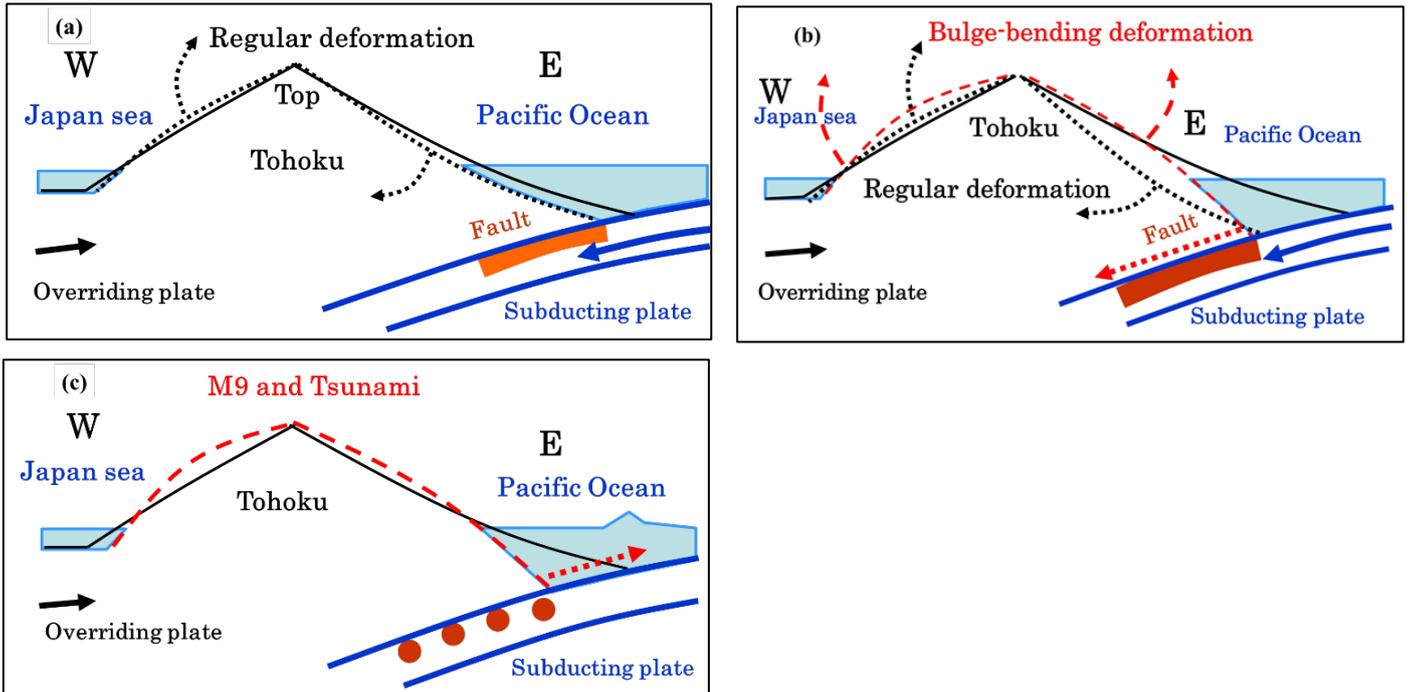

**Figure 4 Three phases of crustal deformation**
Figure 4 is modified and reproduced from references [6, 8]. The labels E, W, and Top correspond to the East coast (Onagawa station), West coast (Ryoutsu2 station), and Top (Murakami station). The east and west coasts of the Tohoku region face the Pacific Ocean and the Sea of Japan, respectively. The Top is the location of the no-displacement-ridge line in Fig. 3. (a) A regular slow deformation began changing the west coast deformation from non-elastic to elastic, generating a restoring force of the compressed west coast with gradual subsidence of 1 to 2.8 mm across Tohoku starting in January 2010. (b) The transition was from the regular to a bulge-bending deformation, which pulled the subducting Pacific Plate by coupling with the fault. As indicated by the eastward arrow, the overriding plate-driving force compressed the elastic west coast, causing the east coast to pull down the subducting plate, as indicated by a dotted arrow on the fault. The east coast's pulling action with the westward displacement began in July 2010. (c) The final phase began in November 2010 with the gradual upheaval growth of 1.2 mm on the east coast, generating the lifting force along the Tohoku subduction zone. The buildup of lifting force on the east coast eventually caused the shear stress to exceed the static frictional strength of the subduction interface weakened by the lifting force, causing the overriding and subducting plates to decouple. The lifting force on the entire Tohoku region helped the decoupling process, releasing a massive recoil force of the compressed west coast against the eastward-plate-driving force. The recoil rapidly restored the bulge-bent deformation of the east and west coasts elastically compressed by the overriding plate-driving eastward force. This rapid restoration led to the Tohoku M9 earthquake and tsunami on March 11, 2011.

**5.2.2.4 Transitional phase (June 2010 to November 2010)**
In June 2010, the transitional phase began with a further gradual subsidence of 3.3 mm on the east coast until the final phase, as illustrated in Fig. 4b. The subsidence firmly grasped the fault and started to pull down the subducting Pacific Plate on July 11, 2010, represented by the dotted arrow over the fault. The pulling action, caused by compressing the west coast, accelerated the westward movement, reaching the highest speed of 0.69 mm/day on December 22, 2010.
About a month before reaching the highest speed, the east coast bulge shifted to the final phase of upheaval growth. The linear upheaval growth of 1.2 mm over 115 days began on November 14, 2010.
The Top began the upheaval growth on August 26, 2010, reaching 2.5 mm before the 2011 M9 EQ event. The gradual upheaval growth of 2.3 mm on the west coast began on October 29, 2010.

**5.2.2.5 Final phase (November 2010 to March 2011)**
In November 2010, the final phase began with the gradual upheaval growth of 1.2 mm on the east coast. The bulge-bending force with this growth decelerated the subducting Pacific Plate's westward movement, which halted by February 21, 2011, and remained motionless for another four days. On February 25, 2011, the bulge force reversed the westward plate motion, with the eastward speed reaching 0.06 mm/day just three days before the M9 EQ on March 11, 2011.

The final upheaval growth of 1.2 mm rapidly released a massive restoring force in the west coast, compressed elastically by the overriding plate-driving eastward force. This recoil restored the bulge-bent west and east coasts, ultimately decoupling the overriding and subducting plates and leading to the megathrust EQ and tsunami on March 11, 2011, as depicted by the dotted arrow in Fig. 4c. Throughout the transitional and final phases, Tohoku experienced an elastic compression of 10.0 mm on the west coast and a pulling movement of 13.2 mm on the east coast in the same westward direction as the subducting oceanic plate-driving westward force, persisting until the 2011 Tohoku EQ event on March 11, 2011.

### 5.2.2.6 Tsunami genesis process

As discussed in the reference [8], the relationship between averaged accelerations (forces) and displacements on the Tohoku subduction zone indicates that the compression of the west coast by the overriding plate-driving eastward force is elastic. The plate-driving eastward force simultaneously compressed (bulge-bent) the east coast westward across Tohoku, not the subducting plate-driving westward force. Thus, the Tohoku M9 earthquake and tsunami were not caused by the commonly suggested elastic rebound of the east coast compressed by the subducting plate-driving force coupled with the over-riding plate through the fault.

The overriding plate-driving eastward force compressed the west coast by +18.4 mm (eastward until the Tohoku M9 EQ on March 11, 2011), initiating the underlying bulge-bending deformation during the initial phase and generating the east coast's pulling action during the transitional and final phases with -13.2 mm (westward) movement. During the final phase, the upheaval growth of 1.2 mm resulted in the accumulation of lifting force along the east coast, which weakened the static frictional strength of the subduction interface. This weakening eventually caused the shear stress to exceed the frictional strength, decoupling the overriding and subducting plates. The decoupling triggered the sudden release of a massive elastic potential energy stored in the compressed west coast, resulting in the rapid restoration of the bulge-bent west and east coast deformation to the regular as an elastic rebound of the entire Tohoku. This entire elastic rebound generated the megathrust EQ and tsunami, as illustrated in Fig. 4c.

### 5.2.2.7 GPS observations consistent with micro-gravity anomaly observations

The continental plate-driving eastward force compressed the west coast eastward and simultaneously caused the east coast to bulge-bend, initiating the process of pulling down the subducting oceanic plate in July 2010. The plate-driving force continued to bulge-bend the east coast until the halt of pulling action in February 2011. The timeline of this pulling action by the enormous continental plate-driving force across Tohoku is consistent with a pre-seismic microgravity anomaly observed by the GRACE satellite data from July 2010 to February 2011 [42]. The upheaval growth that began in November 2010 (characterized by S2 in Fig. 5b) near the Tohoku M9 epicenter area has a timeline that agrees with the uplift growth suggested by the sea-surface gravity change observation from November 2010 to February 2011 [43].

### 5.2.3 Bulge-bending pulling of subducting oceanic plate by the east coast

The daily displacement $d(h, j)$ in Fig. 5 has a background noise of about ± 20 mm. Thus, only the equations of motion with a smooth $D(h, \tau) – V(h, \tau)$ path with the parameters $w = 200$ and $s = 300$ in days can quantify the bulge-bending deformation with subsidence and upheaval by minimizing yearly and seasonal variations and environmental noises in $\{c\}$ ($c = h$). The path has a resolution of 0.1 mm, four orders of magnitude greater than the daily noise level of ± 20 mm. The $D(h, \tau) – V(h, \tau)$ path and $A(h, \tau)$ require displacement $d(h, j)$ ranging from $d(h, \tau − 350)$ to $d(h, \tau + 350)$ and from $d(h, \tau − 500)$ to $d(h, \tau + 500)$, as shown in Eqs. (2) – (4) [8].

The $D(h, \tau) – V(h, \tau)$ path in Fig. 5b shows three approximately linear segments: the initial phase (denoted as S0) with a nearly zero slope, the transitional phase (denoted as S1) with a minus slope in subsidence, and the final phase (denoted as S2) with a positive slope in upheaval growth. All segments experience positive $A(h, \tau)$ that grew linearly in time $\tau$, indicating the bulge-bending deformation on the east coast GPS station was generating a lifting force.

We detail phases S1 and S2 to describe a lifting force generated by the bulge-bending deformation. Each phase showed an approximately linear path on the $D(h, \tau) – V(h, \tau)$ plane, where $V(h, \tau) \approx k1 \times D(h, \tau) + k2$, and $k1$ and $k2$ are constants, and $D(h, \tau)$ is a displacement from an offset origin. The time rate of $V(h, \tau)$ is $A(h, \tau) \approx k1 \times V(h, \tau) \approx k1^2 \times D(h, \tau) + k1 \times k2$. The observed constant $k1$ and $k2$ had the same sign in S1 and S2. The bulge-bending deformations in S1 and S2 show that the $A(h, \tau)$ was always positive. Therefore, it follows that $k1^2 \times D(h, \tau) > − k1 \times k2$, indicating that the positive acceleration (a lifting force on the fault surface) can occur with subsidence of negative $D(h, \tau)$. The lifting force with negative $D(h, \tau)$ is a restoring force. In contrast, a positive $D(h, \tau)$ indicates an upheaval growth caused by bulging, which is always a non-restoring (non-elastic) lifting force. Across the Tohoku region from the west coast to the east coast, the GPS observed the same lifting force.

The positive acceleration is due to bulging that generated a lifting force on the fault surface. The relative force, $F(h, \tau)$, is approximately equal to $A(h, \tau) − k1 \times k2$, which is in turn approximately equal to $k1^2 \times D(h, \tau)$, and indicates a non-restoring upheaval force with positive $D(h, \tau)$ moved up by the bulge-bending deformation. The buildup of lifting force on the east coast gradually reduced the static frictional strength of the subduction interface, eventually causing the shear stress to exceed the frictional strength and leading to the decoupling of the overriding and subducting plates.

Figure 5b shows that segment S1 of the $D(h, \tau) - V(h, \tau)$ path has $\Delta D(h, \tau) = -3.3$ mm (subsidence) and $\Delta V(h, \tau) = +0.0142$ mm/day over 158 days in time $\tau$. It is from $\tau = 3457$ on 24 Jun 2009 to $\tau = 3615$ on 29 Nov 2009, and from $j = 3807$ on 9 Jun 2010 to $j = 3965$ on 14 Nov 2010, in real-time $j$. The average acceleration was $+8.99 \times 10^{-5}$ mm/day$^2$, exerting the upward acceleration (Force) on the east coast GPS station. The average force is indicative of the bulge-bending strength.

Segment S2 has $\Delta D(h, \tau) = 1.2$ mm (upheaval) and $\Delta V(h, \tau) = +0.0054$ mm/day over 115 days from $j = 3965$ on 14 Nov 2010 to $j = 4080$ on 10 Mar 2011, one day before the M9 event. The average acceleration is $+4.7 \times 10^{-5}$ mm/day$^2$.

The $D(h, \tau) - V(h, \tau)$ paths, in Fig. 5, show an approximately constant downward speed of $V(h, \tau) \approx -0.018$ mm/day (Fig. 5b) with $w = 200$ and $s = 300$ except for a significant co-seismic downward displacement. No appreciable change in $V(h, \tau)$ suggests $A(h, \tau) \approx 0$. The $A(h, \tau)$, in Fig. 5b, shows a slight fluctuation until it begins a steady increase at S0. Thus, the regular subsidence deformation at the east coast in Fig. 4a had no significant vertical force component.

Figures 5b and 5c show the positive acceleration (lifting force) in segments S0, S1, and S2, suggesting that they are part of the bulge-bending deformation under the negatively (westwardly) increasing $V(E, \tau)$ and $A(E, \tau)$ with the westward displacement. Thus, the well-known elastic-rebound theory cannot explain the non-restoring force acting on the east coast and the sudden appearance of the upward force component ($A(h, \tau) > 0$) normal to the fault line (surface) shown in Fig.4b.

The downward (westward) bend on the $D(E, \tau) - V(E, \tau)$ path in Fig. 5c started at $\tau = 3434$ on 1 Jun 2009, which is 17 May 2010, in real-time $j = \tau + 350$. The westward bend-onset denoted by Ph-0 preceded S1-onset by 23 days. The westward movement from Ph-0 was $D(E, \tau) = -13.2$ mm by 10 Mar 2011, one day before the Tohoku M9 event.

Segment S1 on the $D(h, \tau) - V(h, \tau)$ path is part of the bulge-bending deformation process at Onagawa station. However, we refer to the transition from S1 to S2 as the bulge-bending deformation onset (bulge-onset) or 'Bulge starts' for the automated detection by real-time power monitoring in section 6.

Segment S1 has a lifting force twice stronger as the S2 force. The S1-onset was at time $\tau = 3457$ (on 14 Jun 2009), including $d(h, j)$ at time $j = 3807$ (on 9 Jun 2010). The date precedes 11 Jul 2010, when the westward motion's trend change began. The S1 ends at the bulge-onset, $\tau = 3615$, 29 Nov 2009, 350 days behind 14 Nov 2010, in real-time $j$. The date precedes 22 Dec 2010, when the abnormal westward speed of the subducting Pacific Plate became about three times higher than the standard speed. Thus, the bulge-bending deformation is the geophysical origin of the abnormal westward motion of the subducting oceanic plate.

We note that each linearity of S1 and S2 holds for the respective $D(h, \tau) - A(h, \tau)$ path. However, $A(h, \tau)$ at the S2-onset ($\tau = 3615$) requires $d(h, j)$ at $j = 4115$ ($> j = 4080$, 10 Mar 2011). Thus, the second linear segment is not available.

As in Figs. 6d and 6g, the observation in low-frequency selection accompanies some onset-detection delay in real-time $j$. However, the equations with $w \approx 200$ and $s \approx 300$, even with a much longer real-time delay, are requisite for quantifying the Tohoku's bulge-bending deformation in a three-phase process by minimizing the yearly and seasonal variations and environmental noises in $\{c\}$ ($c = h$).

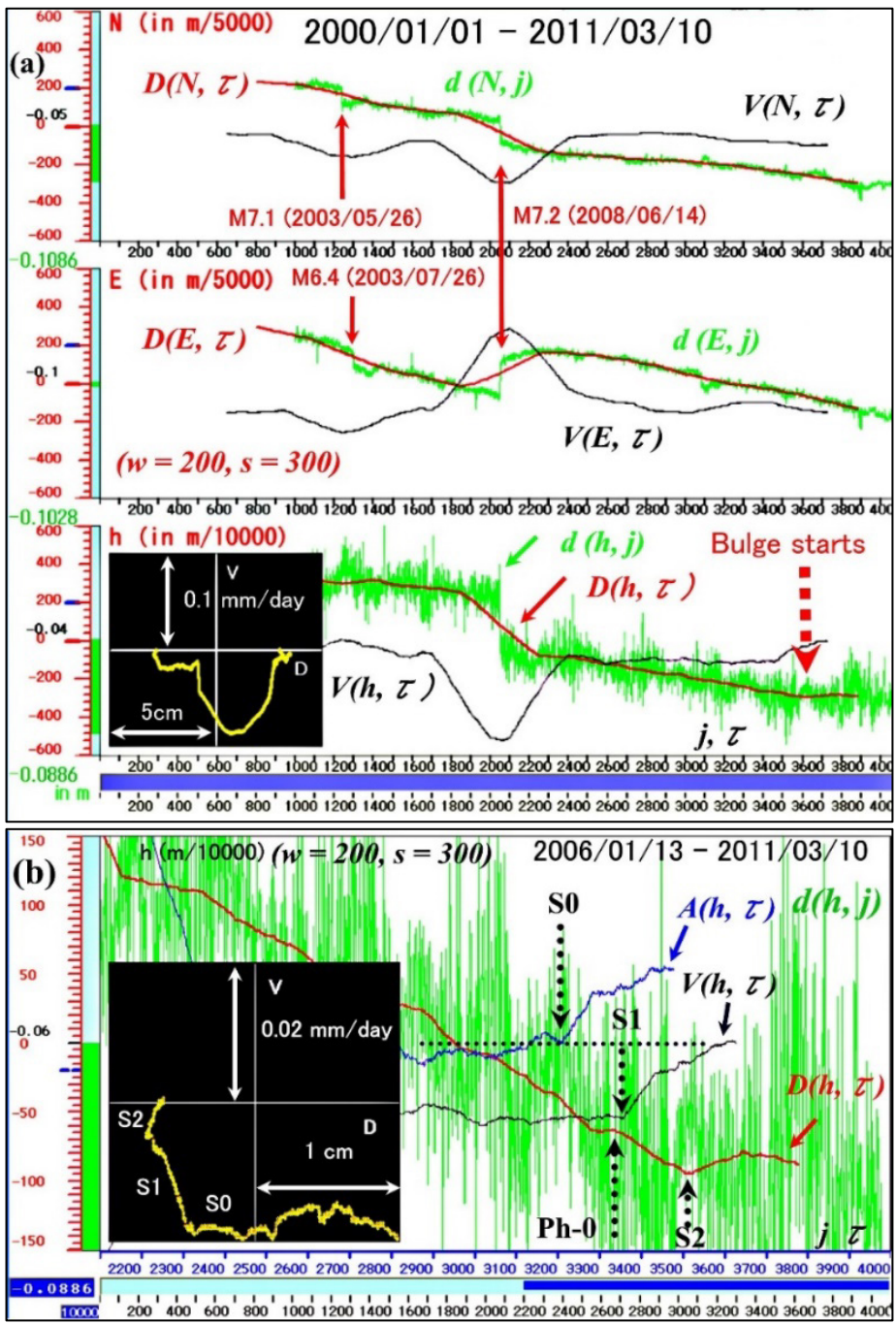

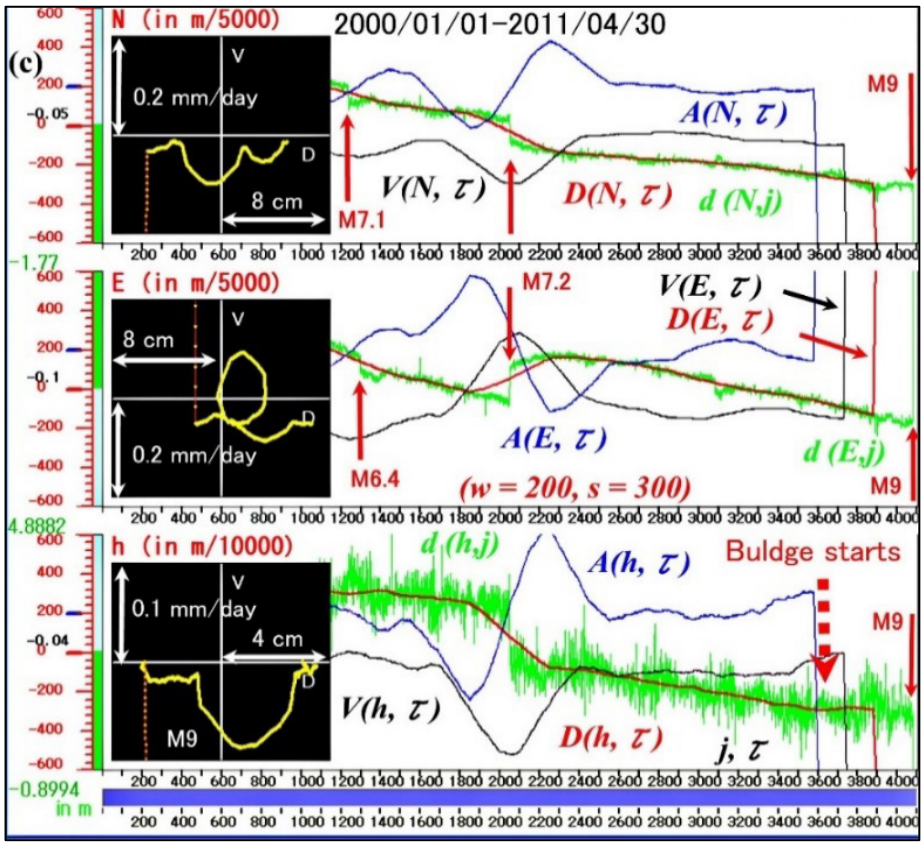

**Figure 5. The bulge and M9 observations at Onagawa station (east coast)**
Figure 5 is modified and reproduced from references [6, 8]. (a) The series {c} covers the period from January 1, 2000 (j = 0) to March 10, 2011 (j = 4080), with sudden large co-seismic shifts. The magnitudes and dates are indicated on d (c, j) at arrows. Parameters w = 200 and s = 300 are used for D (c, τ), V (c, τ), and A (c, τ). The D (h, τ) – V (h, τ) phase plane has a scale of (5 cm, 0.1 mm/day) with the offset origin D (h, τ) = − 0.04 m and V (h, τ) = 0 mm/day. The bulge-onset is labeled at arrow 'Bulge starts' on the D (h, τ) at τ = 3615, November 29, 2009 (November 14, 2010, in time j). (b) The expanded window is from January 13, 2006, to March 10, 2011. The magnified phase plane has a scale of (1 cm, 0.02 mm/day) with the origin D (h, τ) = − 0.062 m (− 0.06 m − 0.002 m). Approximately linear segments are labeled as S0, S1, and S2 on the phase path. Each segment's onset location is at a dot-arrow with its name. The S2 is the 'Bulge starts' in Fig. part a. Arrow Ph-0 between S0 and S1 is the onset location of an abnormal movement on the D (E, τ) – V (E, τ) path in Fig. part c. A horizontal dot-line at the offset origin 0 (− 0.06) is the abscissa for V (h, τ) and A (h, τ); namely, V (h, τ) = A (h, τ) = 0. The magnitudes of V (h, τ) and A (h, τ) are on relative scales. (c) The series {c} covers the period from January 1, 2000 (j = 0) to April 30, 2011 (j = 4107). It includes the M9 EQ on March 11, 2011 (j = 4081), arrow M9 on d (c, j). The D (c, τ) – V (c, τ) planes have a scale of (8 cm, 0.2 mm/day) with the offset origins for c = N and E, and (4 cm, 0.1 mm/day) for c = h. The D (c, τ) – V (c, τ) path shows the doted path jumped by the M9 EQ. Every d (c, j) and its column height have the co-seismic shifts saturated by the M9 EQ with its digital value from the position at j = 0. The southward shift is 1.90 m (− 1.77 − 0. 05 − 0.08), the eastward shift is 4.7482 m (4.8882 − 0. 10 − 0. 04), and the downward shift is 0.9694 m (= − 0.8994 − 0. 04 − 0. 03).

### 5.2.4 Abnormal movements of the subducting plate pulled by the bulge-bending east coast

The oceanic plate is coupled with the overriding Tohoku crust through the fault, as illustrated in Fig. 4c. There are three GPS stations on the oceanic plate's islands, as shown in Fig. 3. These GPS stations provide daily displacement time series {c}, representing the oceanic plate motion, which is under the periodic lunar synodic tidal force loading. Monitoring the periodic lunar responses of the oceanic plate motion to the bulge-bending pulling by the overriding eastern edge is crucial for detecting subtle changes and abnormal movements that may lead to a megathrust EQ and tsunami.

To monitor the periodic lunar responses of the oceanic plate motion, we define the displacement $D(c, \tau)$, velocity $V(c, \tau)$, and acceleration $A(c, \tau)$ as Eqs. (2) – (4) with w = 7 and s = 20 for a period of 30 (29.5) days.

As seen in Fig. 6 for the Chichijima station, an unexpectedly increased westward speed appeared on the $D(E, \tau)$ – $V(E, \tau)$ path. A trend-change on $D(E, \tau)$ in Jul 2010 preceded the abnormal motion whose westward speed $V(E, \tau)$ reached its highest on 22 Dec 2010, 76 days before the Tohoku M9 EQ, at − 0.69 mm/day, approximately three times higher than the standard westward speeds. A rapid deceleration followed, stopping the westward motion around 21 Feb 2011. In about four days, the moving direction reversed, and the subducting plate moved eastward until the Tohoku M9 events on 11 Mar 2011. Other GPS stations in the Northwest Pacific Ocean observed the same abnormal movements as in Table 1 [8].

**Table 1 (Abnormal motion of the subducting northwestern Pacific Plate)**

| GPS station | $D(E,\tau)$ trend change | | Max $V(E,\tau)$ | | $V(E,\tau)$ and $D(E,\tau)$ mm on March 8 and 10 | | | |
|---|---|---|---|---|---|---|---|---|
| | yyyy/mm/dd | $V(E,\tau)$ | yyyy/mm/dd | $V(E,\tau)$ | March 8 | | March 10 | |
| Chichijima | 2010/07/11 | −0.25 | 2010/12/22 | −0.69 | +0.06 | +1.6 mm | | |
| | No observation after 2011/03/08 | | | | | | | |
| Chichijima-A | 2010/07/11 | −0.26 | 2010/12/22 | −0.78 | +0.08 | +2.0 mm | +0.12 | +2.4 mm |
| Hahajima | 2010/07/11 | −0.28 | 2010/12/22 | −0.84 | +0.12 | +2.0 mm | +0.20 | +2.5 mm |
| Minamitorishima | 2010/07/11 | −0.34 | 2010/10/08 | −1.15 | 0.00 | 0.0 mm | +0.05 | 0.0 mm |
| | No data available; 2010/08/07〜2010/09/25, 2010/10/13〜2011/01/01 | | | | | | | |

The power monitoring described in section 6 detected the onset of the trend-change and the abnormal motion, as shown in Figs. 6d and 6e. The lunar synodic loading can be removed with $w = 15$ and $s = 40$, as shown in Fig. 6f. The $PW(E,j)$ monitoring detected the anomaly onset more clearly, as shown in Figs. 6g and 6h. The unexpected trend-change on $D(E, \tau)$ was in July 2010 in Fig. 6h, which was the first response of the subducting oceanic plate to the bulge-bending deformation of the overriding eastern edge.

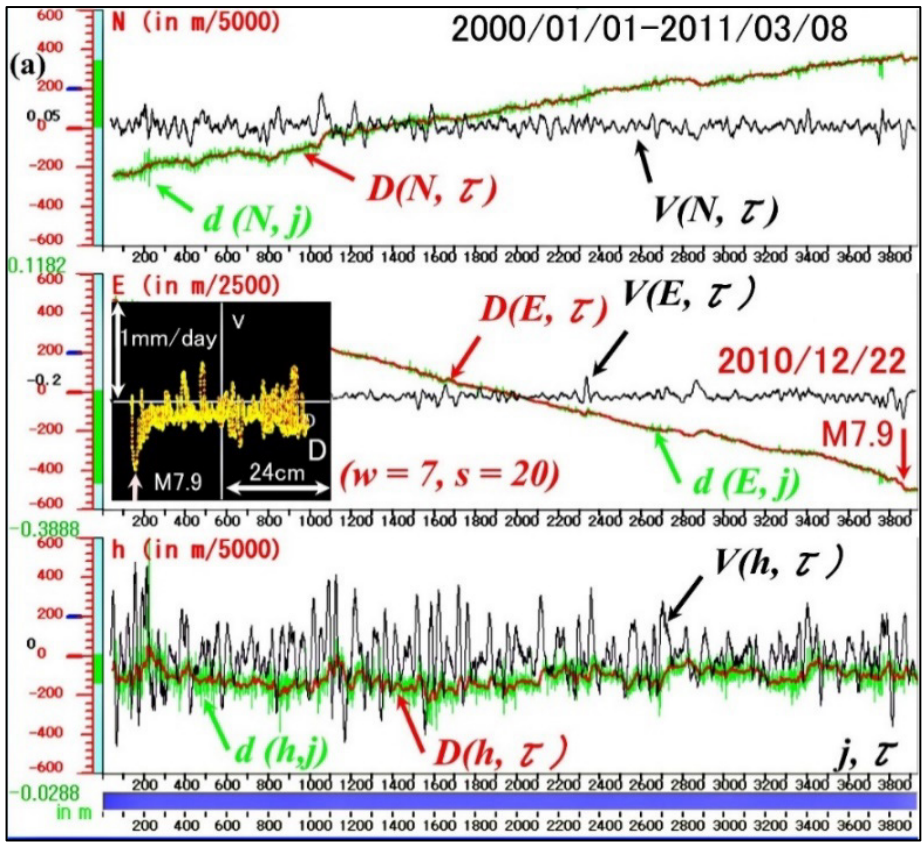

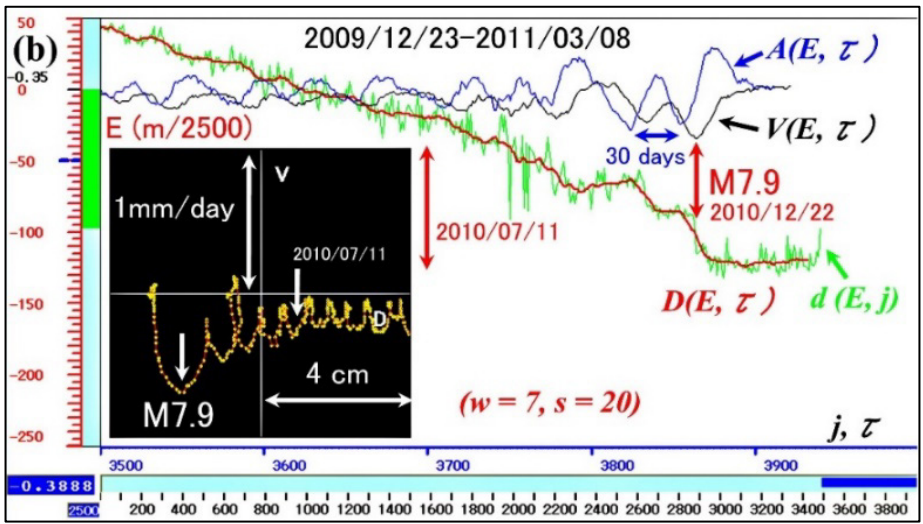
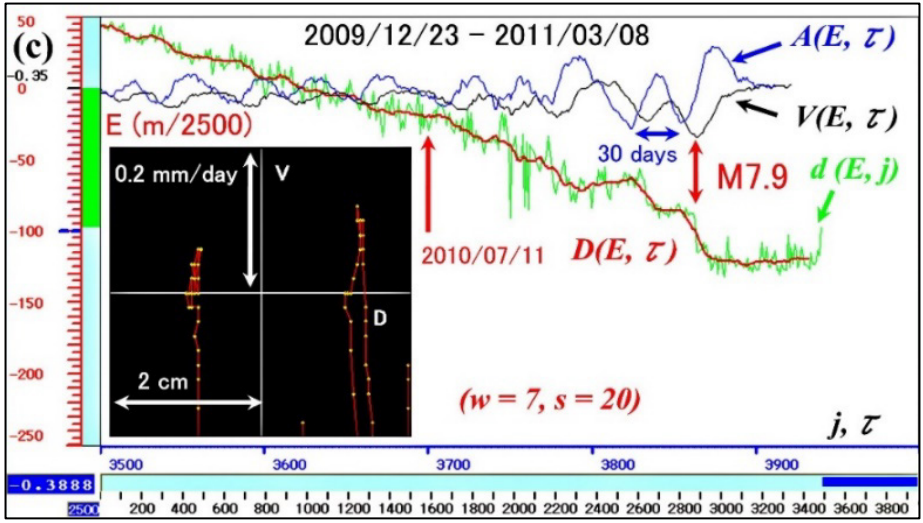
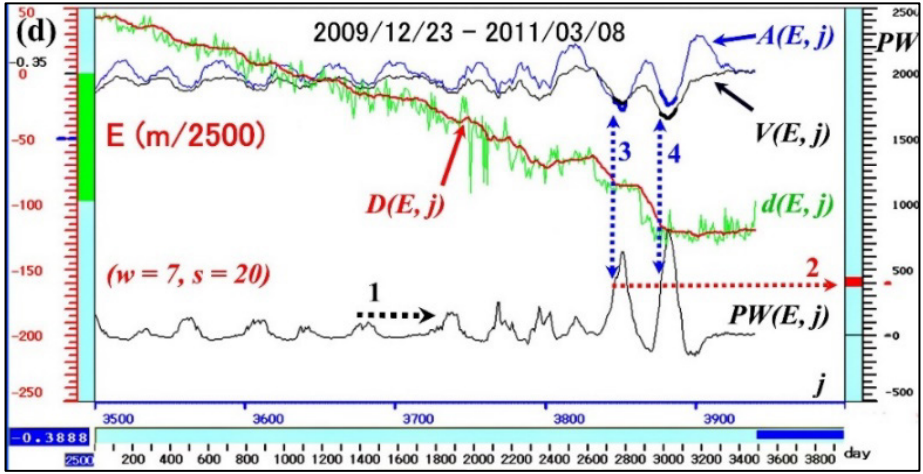

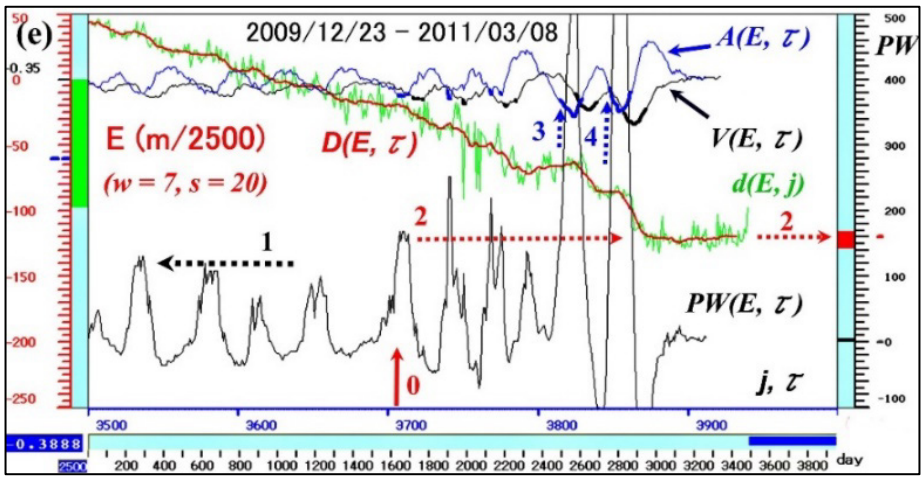
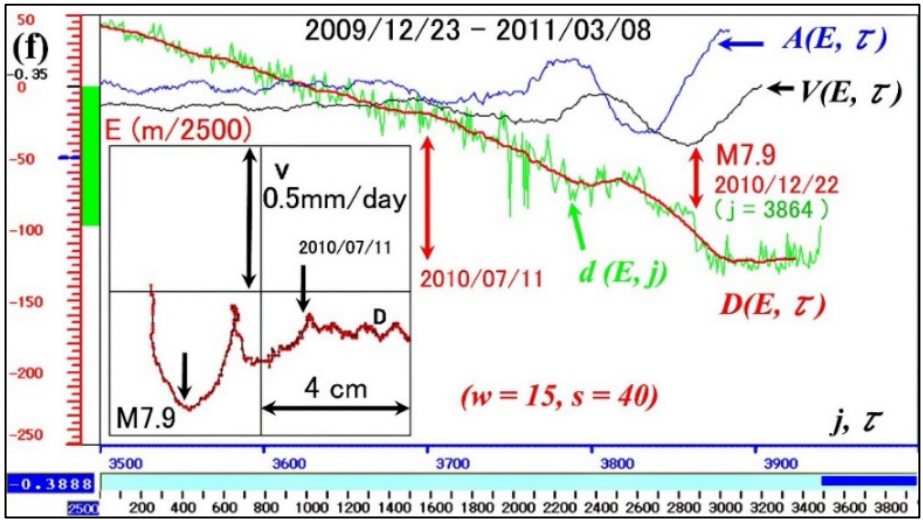
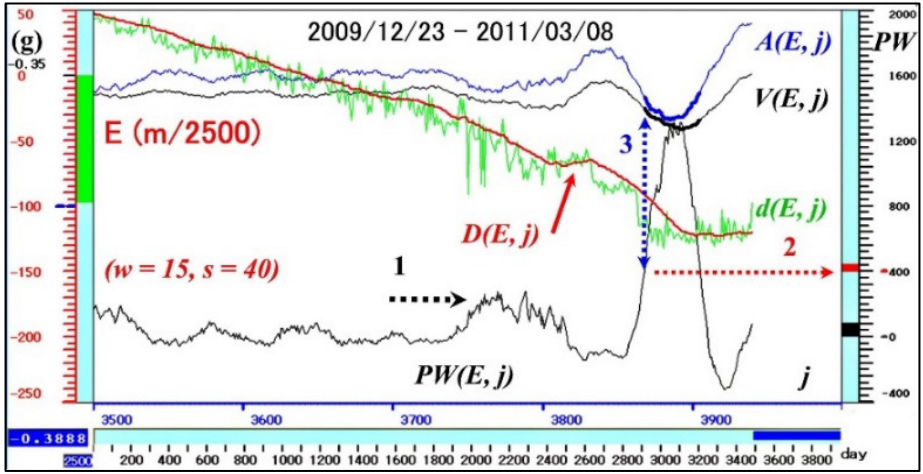

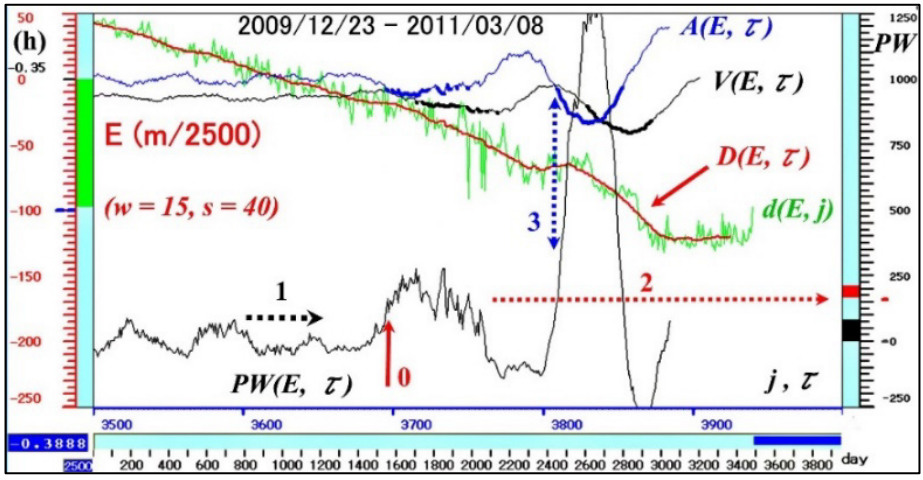

**Figure 6. The abnormal westward motion at Chichijima station**
Figure 6 is modified and reproduced from references [6, 8]. (a) The series {c} covers the period from $j$ = 0 (January 1, 2000) to $j$ = 3940 (March 8, 2011), with parameters $w$ = 7 and $s$ = 20 used for $D (c, \tau)$, $V (c, \tau)$, and $A (c, \tau)$. The $D (E, \tau) - V (E, \tau)$ plane has a scale of (24 cm, 1 mm/day), with an origin at $D (E, \tau) = -0.2$ m, which is the offset reference (− 0.2 m) at scale 0 from the first-day position ($j$ = 0). The $V (E, \tau)$ origin is 0 mm/day. The right half of the $D (E, \tau) - V (E, \tau)$ plane is east of the offset origin. The upper half is eastward and positive $V (E, \tau)$, while the lower half is westward and negative. The M7.9 EQ (2010/12/22) is located on $D (E, \tau)$ at $\tau$ = 3864, and the (24 cm, 1 mm/day) path. The $d (c, j)$ is in green, $D (c, \tau)$ in red, and $V (c, \tau)$ in black. The $V (c, \tau)$ is in a relative scale from the graphical origin 0. (b) The expanded time window is from $j$ = 3500 (December 23, 2009) to $j$ = 3940 (March 8, 2011). The (4 cm, 1 mm/day) phase plane has an offset origin at the blue-line-scale − 50, $D (E, \tau)$ = − 0.22 m (− 0.2 − 50/2500). The $V (E, \tau)$ and $A (E, \tau)$ are in relative scales from the same graphical origin 0. The blue $A (E, \tau)$ with a period of 30 days represents the lunar synodic tidal force loading (29.5-day period). The date label 2010/07/11 corresponds to the $D (E, \tau)$ trend change at $\tau$ = 3700 (July 11, 2010). (c) The window is the same as Fig. part b, with a magnified (2 cm, 0.2 mm/day) plane having the offset origin at the blue-line-scale −100, $D (E, \tau) = -0.39$ m (− 0.35 m − 100 × m/2500). The path ends at $V (E, \tau)$ = + 0.06 mm/day. (d) The window is the same as Fig. part b with a $PW (E, j)$ monitoring. The relative power scales are on the right column. The monitoring with $PW (E, j) \geq$ 400 detected an anomalous lunar synodic loading on {$E$}. The predetermined 400 was about twice the expected standard power level at arrow 1. Level 400 is at the red scale at arrow 2. Arrow 3 was the first anomaly detection at $j$ = 3847 (December 5, 2010). At the detection, $PW (E, j)$ rose from 378 (at $j$ = 3846) to 440 (at $j$ = 3847), showing the red column height. Arrow 4 was the second detection at $j$ = 3877 (January 4, 2011). The anomalous $V (E, j)$ and $A (E, j)$ were highlighted in bold under $PW (E, j) \geq$ 400. (e) The window is the same as Fig. part b with $PW (E, j) \geq$ 160, finding the westward trend change at time $j$ = 3735 on August 15, 2010. In time $\tau$, the change was at $\tau$ = 3708 (at arrow 0) on July 19, 2010. The detecting level 160 at arrow 2 was adopted from the standard power level at arrow 1. By shifting time $j$ back to time $\tau$, the displays are $D (E, \tau)$, $V (E, \tau)$, $A (E, \tau)$, and $PW (E, \tau)$. (f) The time window is the same as Fig. part b, with parameters $w$ = 15 and $s$ = 40, removing the 30-day-period oscillation. The (4 cm, 0.5 mm/day) plane has the offset origin, $D (E, \tau) = -0.37$ m (− 0.35 − 0.02) at the blue-line scale − 50. The $D (E, \tau)$ trend change has the date label 2010/7/11. (g) The window is the same as Fig. part b with $w$ = 15 and $s$ = 40. Arrow 3 was the anomaly detection at $j$ = 3869 (December 27, 2010) by $PW (E, j) \geq$ 400. The threshold 400 is at the red scale pointed by arrow 2. The threshold was adopted from the standard power level at arrow 1. The power column height change in red is from 400 (at $j$ = 3868) to 442 (at $j$ = 3869) at the anomaly detection. The black column height shows the last $PW (E, j)$ (just above level 0) on March 8, 2011 ($j$ = 3940). (h) The window is the same as Fig. part b for the $D (E, \tau)$ trend change detection. Up-arrow 0 was the trend change detection at time $j$ = 3751 (August 31, 2010) by $PW (E, j) \geq$ 160 with $w$ = 15 and $s$ = 40. In time $\tau$, the change was at arrow 0, $\tau$ = 3696 on July 7, 2010. Arrow 1 is the standard power level, for which the unexpected level 160 was about twice the standard level.

As in Fig. 6a, series {$E$} in the second window shows the $D(E, \tau)$ and $V(E, \tau)$ relation expressed as a path on the (24 cm, 1 mm/day) plane. At the highest westward (downward) speed, the M7.9 EQ ruptured in the Pacific about 187 km away from the station (Fig. 3). The event had normal faulting (Strike, Dip, Rake) = (340º, 57 ˚, − 56 ˚) [25], suggesting the abnormal westward motion triggered the event.

In Fig. 6b, the $D(E, \tau)$, $V(E, \tau)$, and $A(E, \tau)$ show the lunar synodic tidal force loading (the 29.5-day-period) on the subducting Pacific Plate (Chichijima station). The path moves eastward, protruding by the amount of 0.2 mm/day from the westward speed at − 0.23 mm/day. Dividing the 6 mm separation between the two protruding peaks by 30 days, the westward speed estimation is − 0.2 mm/day, which qualitatively shows the periodic oscillation is due to the synodic tidal loading. The westward trend of $D (E, \tau)$ changed at around $\tau$ = 3700 (July 11, 2010), indicated by the labeled arrow, 2010/07/11, on the $D (E, \tau)$. The trend change onset occurred due to an insufficient synodic tidal loading on the roughly linear segments of $D (E, \tau)$, $V (E, \tau)$, and $A (E, \tau)$. The onset divided the $D (E, \tau) - V (E, \tau)$ path into two small linear segments. The first linear segment preceding the trend change has an explicit segment equation $A(E, \tau) \approx K \times D(E, \tau)$, as in the $D(E, \tau) - A(E, \tau)$ path segment of Fig. 7. The constant $K$ is positive. However, $K$ is negative under effective lunar synodic tidal loading, obeying the oscillatory motion. Following the trend-change, the motion became anomalous and reached the highest westward speed of $V(E, \tau) = -0.69$ mm/day at $\tau = 3864$ on 22 Dec 2010, approximately three times faster than $V(E, \tau) = -0.25$ mm/day at $\tau = 3700$ on 11 Jul 2010. After this, the westward motion showed a rapid deceleration until it stopped at $\tau = 3908$ on 4 Feb 2011.

Figure 6c displays a magnified path, changing $V(E, \tau)$ positive on 8 Feb 2011. The reversed motion reached the eastward speed of $V(E, \tau) = +0.06$ mm/day and the eastward displacement of 1.6 mm at time $\tau = 3918$ on 14 Feb 2011. In real-time $j$, it was 8 Mar 2011, three days before the 11 Mar M9 EQ, because $j$ is in advance of $\tau$ by 17 days ($\tau = j - s/2 - w$).

Figure 6d shows the $PW(E, j) \geq 400$ monitoring detected the anomaly by a predetermined threshold level of 400. Level 400 is an unexpected power level twice the standard $PW(E, j)$ amplitudes at dot-arrow 1. The threshold adoption can be automatic or manual during power monitoring. Arrow 3 and Arrow 4 were the first at $j = 3847$ (5 Dec 2010) and second at $j = 3877$ (4 Jan 2011) anomaly detections, respectively. At each detection, $V(E, j)$ and $A(E, j)$ become bold, and they remain bold as long as $PW(E, j) \geq 400$. They are negative, indicating the abnormal movement was westward.

Figure 6e shows that the unexpected westward motion started at the upward arrow 0. The $PW(E, j) \geq 160$ monitoring found the westward trend-change at real-time $j = 3735$ on 15 Aug 2010. In time $\tau$ ($\tau = j - s - w$), it was 3708 on 19 Jul 2010. We shifted real-time $j$ back to time $\tau$ to draw $D(E, \tau)$, $V(E, \tau)$, $A(E, \tau)$, and $PW(E, \tau)$, satisfying each time reversal property. As for anomalies 3 and 4, the negative $A(E, \tau)$ precedes negative $V(E, \tau)$ so that the anomalous acceleration (oceanic tectonic driving force) and the motion to follow are westward.

In Fig. 6f, we define the $D(E, \tau)$, $V(E, \tau)$, and $A(E, \tau)$ with $w = 15$ and $s = 40$, which masked the lunar synodic loading. The $D(E, \tau) - V(E, \tau)$ path shows the abnormal $V(E, \tau)$ more apparent than that in Fig. 6d.

Figure 6g shows that the $PW(E, j) \geq 400$ monitoring with $w = 15$ and $s = 40$ detected the abnormal motion at arrow 3 (at $j = 3869$ on 27 Dec 2010). The power change is the red column height. The $V(E, j)$ and $A(E, j)$ in bold are negative so that the abnormal motion was by the westward $A(E, j)$.

Figure 6h shows the $PW(E, \tau) \geq 160$ monitoring with $w = 15$ and $s = 40$ detecting the westward trend change without the effective synodic tidal force loading on $D(E, \tau)$. The trend change is independent of the lunar tidal force loading, supporting that the geophysical origin of the trend change was a transition from regular subsidence deformation on the east coast of Tohoku to bulge-bending deformation, as explained in section 5.3. The abnormal westward motion of the subducting northwestern Pacific Plate followed the trend change.

The $PW(E, j)$ monitoring with the low-frequency $D(E, \tau)$, $V(E, \tau)$, and $A(E, \tau)$ creates the anomaly-onset-detection delay in time $j$. The delay allowance depends on the detecting objectives. For example, the unexpected $V(E, j)$ and $A(E, j)$ detection time in Fig. 6d was $j = 3847$ (5 Dec 2010), which could be useful to predict the imminent M7.9 event. However, in Fig. 6-1g, with $w = 15$ and $s = 40$, the detection was on 27 Dec 2010 at $j = 3869$, which followed the trend change in $D(E, \tau)$ and was useful for predicting the Tohoku M9 EQ on 11 Mar 2011. Automated power monitoring with multiple frequencies and thresholds for abnormal event detections is always available.

The overriding edge pulled the subducting oceanic plate at the abnormal westward speed of $V(E, \tau) = -0.69$ mm/day, as in Table 1. The $D(E, \tau) - A(E, \tau)$ paths in Figs. 7a – 7d exhibit the lunar synodic tidal force $F(E, \tau)$ loading under the subducting plate's westward uniform movement by a segmented equation, $F(E, \tau) \approx A(E, \tau) \approx K \times D(E, \tau)$ with a constant $K$ (positive or negative). The periodic tidal force loading on the plate motion shows negative $K$ under no other external force loading. A standard tidal force loading with negative $K$ stands for no external force except the tidal force on the plate motion. The standard tidal force must have an appropriate range of $\Delta D(E, \tau)$ and $\Delta A(E, \tau)$ on the $D(E, \tau) - A(E, \tau)$ path like the ranges before the trend change on 2010/07/11 in Figs. 7a – 7d.

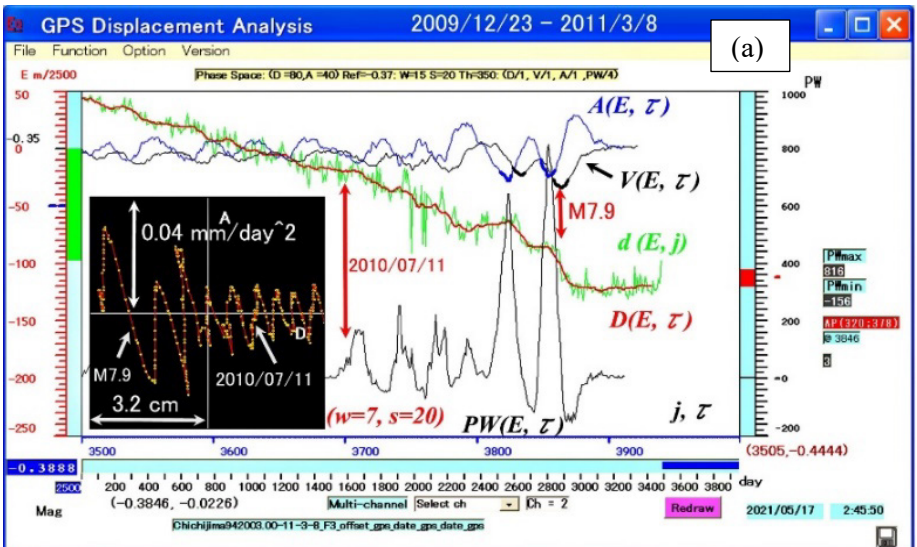
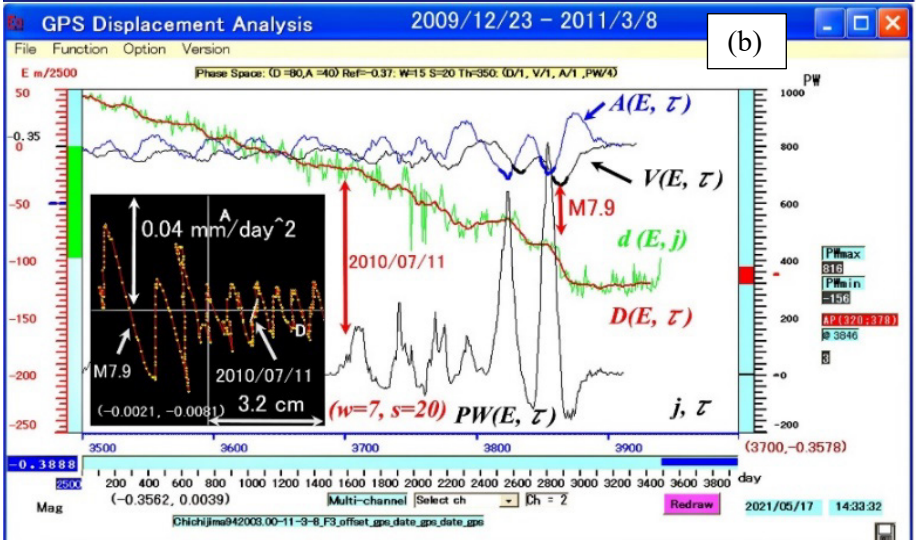
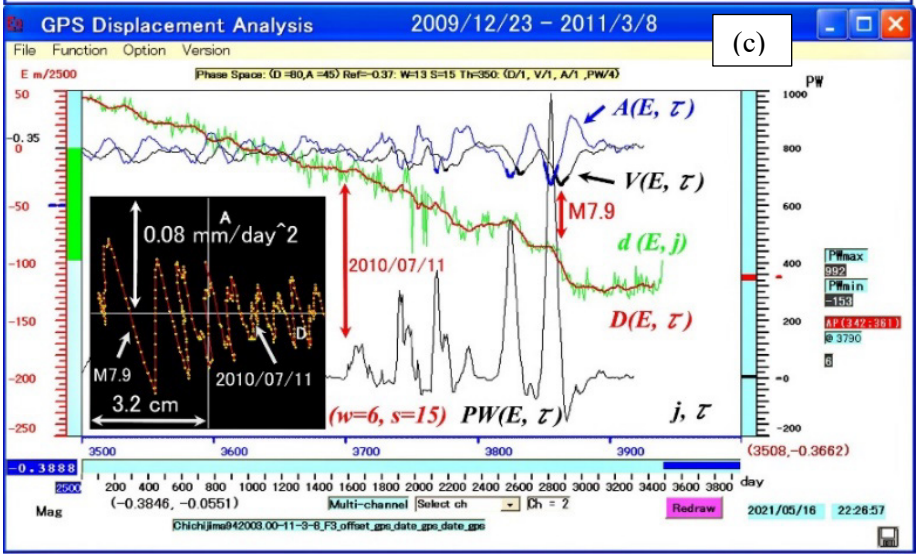

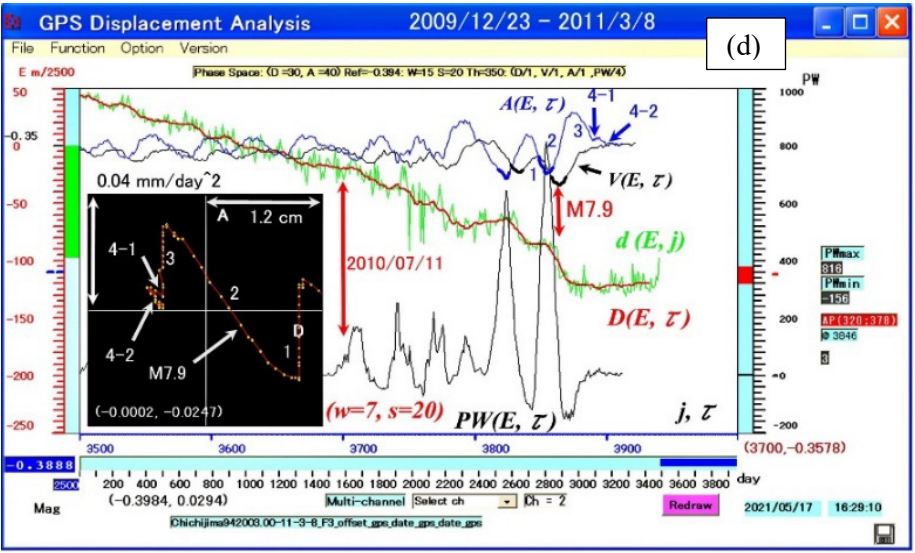

**Figure 7.** $D(E, \tau) - A(E, \tau)$ path for Fig. 6b
Figure 7 is reproduced from reference [8]. (a) The $D(E, \tau) - A(E, \tau)$ path with $w = 7$ and $s = 20$ at Chichijima station for Fig. 6b. (b) The $D(E, \tau) - A(E, \tau)$ path with a segment equation of $A(E, \tau) \approx K \times D(E, \tau)$ for Fig. 6b. The westward trend-change at arrow 2010/07/11 is indicated in a white segment line. (c) The $D(E, \tau) - A(E, \tau)$ path with $w = 6$ and $s = 15$ shows a higher frequency structure of $A(E, \tau)$ in the white segment. (d) A magnified $D(E, \tau) - A(E, \tau)$ path for the last three months before the Tohoku M9.

Figure 7b shows a linear segment of the $D(E, \tau) - A(E, \tau)$ path as a white line. The linear change is ($\Delta D(E, \tau)$, $\Delta A(E, \tau)$) = ($-0.0021, -0.0081$) in (m, mm/day$^2$) from $A(E, \tau) = 0.0039$ mm/day$^2$ at $\tau = 3681$ on 22 Jun 2010, as shown below the time scale ($-0.3562, 0.0039$) in (m, mm/day$^2$). Thus, $A(E, \tau) \approx K \times D(E, \tau)$ has $K = 0.0039$ /day$^2$ because $K = \Delta A(E, \tau) / \Delta D(E, \tau) = 3.85^{-2}$/day$^2$. The (3700, $-0.3578$) in ($\tau$, m) below the PW scale shows the $D(E, \tau)$ reading at the trend change. The segment in a white line in Fig. 7b preceded the trend change at label 2010/07/11. It had a positive $K$ in $A(E, \tau) \approx K \times D(E, \tau)$, suggesting that the standard tidal loading began to couple with the east coast bulge. Thus, $K$, $\Delta D(E, \tau)$, and $\Delta A(E, \tau)$ are quantitative indicators of external force coupling with the standard tidal loading.

Figure 7c shows that an insufficient synodic tidal loading on the segment appears on the $D(E, \tau) - A(E, \tau)$ path in a higher frequency resolution with $w = 6$ and $s = 15$. A fine structure of the white line segment (in Fig. 7b) is a three-segmented $A(E, \tau)$ at arrow 2010/07/11. The first $A(E, \tau)$ changes from up (eastward) to down (westward) with a positive $K$. The second changes opposite the first, from westward to eastward, having a negative $K$. The third $A(E, \tau)$ is the same as the first with a positive $K$. The positive $K$ shows that the plate's westward movement had the standard tidal loading constrained by the overriding Tohoku eastern edge bulge bending. A similar fine structure preceding the trend change shows a repeated constraint by the east coast bulge coupling with the westward movement.

Figure 7d shows four linear segments of the $D(E, \tau) - A(E, \tau)$ path, magnified in indexed order for Fig. 6b. The $A(E, \tau)$ has the corresponding indexes. The path segments are the last part of the M9 EQ genesis process. Each segment has the index number, starting time $\tau$, ($\Delta D(E, \tau)$ mm, $\Delta A(E, \tau)$ mm/day$^2$), and $K$ in $A(E, \tau) \approx K \times D(E, \tau)$. Segment 4 has two sub-portions, 4-1 and 4-2. Last segment 4-2 ends at $\tau = 3913$ on 9 Feb 2011 because $j = 3940$ is 8 Mar 2011 ($\tau = 3940 - w - s$).

| Segment | $\tau$ | Date | ($\Delta D(E, \tau)$ mm, $\Delta A(E, \tau)$ mm/day$^2$) | $K$ (1 /day$^2$) |
|---|---|---|---|---|
| 1 | 3844 | 2 Dec 2010 | (0.1 mm, $-0.0330$ mm/day$^2$) | $K = -0.0330$ /day$^2$ |
| 2 | 3854 | 12 Dec 2010 | ($-8.7$ mm, $0.0434$ mm/day$^2$) | $K = -0.0050$ /day$^2$. |
| 3 | 3878 | 5 Jan 2011 | ($-0.2$ mm, $-0.0247$ mm/day$^2$) | $K = 0.1235$ /day$^2$. |
| 4-1 | 3890 | 17 Jan 2011 | ($-1.6$ mm, $0.0034$ mm/day$^2$) | $K = -0.0213$ /day$^2$. |
| 4-2 | 3894 | 21 Jan 2011 | (1.5 mm, $-0.0068$ mm/day$^2$) | $K = -0.0453$ /day$^2$. |

The oceanic plate has the unusual tidal force coupling of 33 days in segments 1 and 2. Constant $K$ in segment 2 is minimal, indicating no effective tidal force loading. It suggests that the tidal loading had an avalanche-like effect on the plate's westward motion. The bulging eastern edge pulled the plate westward with the highest velocity that triggered the M7.9 at $\tau = 3864$ on 22 Dec 2010, the arrow on the $D(E, \tau) - A(E, \tau)$ path at $A(E, \tau) \approx 0$. Segment 3 then shows the abnormally high external force loading with $K = 0.1235$ /day$^2$ on the standard tidal loading, indicating that the bulge pulling has a sudden deceleration. Segment 4-1 recovers the standard tidal loading for only five days. Segment 4-2 shows the east coast moving eastward by 1.5 mm to prepare the rupturing process of the megathrust EQ, as in Fig. 4c.

# 6 Automatic detection of anomalies

Power monitoring in real-time is crucial for automatically detecting anomalies whose amplitudes are comparable to observed fluctuations in time-series data{c} [8, 33–39]. The power is defined as the time-rate change of the kinetic energy which can be expressed as $PW(c, \tau) = V(c, \tau) \times A(c, \tau)$ with some variations [33]. For the real-time monitoring at the current time $j = \tau + w + s$ in $A(c, \tau)$, the power is $PW(c, j) = V(c, j) \times A(c, j)$. The power at time $j$ is given by:

$$PW(c, j) = V(c, j) \times A(c, j) = \frac{1}{s} KE(c, j) \times \left(1 - \frac{V(c, j-s)}{V(c, j)}\right) . \qquad (5)$$

The $KE(c, j)$ is the kinetic energy, defined as the squared of $V(c, j)$. The velocity $V(c, j)$ captures the fluctuations of period $2s$ where $D(c, j)$ and $D(c, j-s)$ will have the opposite sign. The $KE(c, j)$ magnifies the relative velocity change in Eq. (5) parentheses. Thus, power $PW(c, j)$ becomes maximum near either troughs or peaks of the periodic fluctuations of $2s$ in $A(c, j)$. The larger the amplitude of $A(c, j)$ localized within $2s$ becomes, the larger $PW(c, j)$. Thus, any anomaly becomes the corresponding large $PW(c, j)$. A predetermined threshold detects the rising and falling $PW(c, j)$, which becomes higher than the threshold level and gives the anomaly-onset time. The threshold level may automatically adopt the $PW(c, j)$'s maximum amplitude during the standard condition. A detailed example of power monitoring is in a GPS displacement analysis [8]. $PW(c, j)$ can be applied to automatic anomaly detection by replacing $D(c, \tau)$ with any noisy signal or chaotic time series data observed by physical systems.

# 7 A strain-energy cycle during the significant and megathrust EQ genesis processes

Moving sums of $2s$ $d(INT, j)$s and $2s$ $d(DEP, j)$s in a small or large region are the functions of noise-free principal stress components, which are also the density of strain energy in the region [6, 7]. Significant EQ genesis processes in every mesh of about 4° by 5° in Japan accompany the strain energy accumulation and release cycles [6, 7, 12–15]. Their normalized cycles with the past maximums are $NCI(m, 2s)$ for $2s$ $d(INT, j)$s and $NCD(m, 2s)$ for $2s$ $d(DEP, j)$s. The $NCI(m, 2s)$ is proportional to an averaged $INT$ and inversely proportional to seismic activity. The activity is quiet if the EQ's emerging average rate is slow. The $NCD(m, 2s)$ is proportional to an averaged $DEP$. If it is large, the seismic activity is deep. The $NCD(m, 2s)$ and $NCI(m, 2s)$ are scale-dependent tools to detect the critical stress (strain energy) build-up in the B upper crust and the D-B transition region [7].

The $NCI(m, 2s)$ and $NCD(m, 2s)$ increase during the significant EQ genesis processes. In Fig. 8a, the strain-energy accumulations in $NCI(m, 70)$ and $NCD(m, 70)$ reach the quietest and deepest seismicity at about $m = 556$ (6 Apr 1994), which is $t = 521$ with $m = t + 35$. The energy accumulation peaks around the highest at $t = 528$ in Fig. 8a during CQK for the 1995 Kobe M 7.2. After reaching the peak, a rapid strain energy release into shallower seismicity began and continued until the 1995 Kobe event at $m = 602$ (17 Jan 1995). Figure 8a shows that a new strain-energy accumulation and release process started for the 1997 Yamaguchi M6.7 (CQT) after the 1995 Kobe M7.2, and another new process began for the 2000 Tottori M7.2 (CQT). Thus, a cycle of strain energy accumulation and release repeats, as shown in $NCI(m, 2s)$ and $NCD(m, 2s)$. The so-called Accelerated Moment Release (AMR) increases the background seismicity during the strain-energy release.

The $NCI(m, w)$ and $NCD(m, w)$ as in Fig. 8a can predict rupture date for the large EQ with a high degree of accuracy. These normalized cycles generally peak a few days before a significant event occurs in wide regions. Changes in the regional size, $Mc$, and $w$ can increase the prediction accuracy within a day. For example, Fig. 8b shows the 1995 Kobe M7.2 in the large region of 16°–52° N and 116°–156° E with $Mc = 4$ and $w = 30$. The $NCI(m, 30)$ peaks at $m = 9892$ (14 Jan 1995, at 04:49), and $NCD(m, 30)$ peaks at $m = 9906$ (16 Jan 1995, at 08:53). The Kobe event occurred at $m = 9909$ (17 Jan 1995, at 05:40).

Figures 8a and 8b show that $NCI(m, 2s)$ and $NCD(m, 2s)$ increase to their peaks together, and then they rapidly decrease from their peaks, during which a significant shallow event occurs. If the expected event is significant and deep, like EQs in the Wadati-Benioff zone, $NCD(m, 2s)$ keeps increasing. Figure 9 illustrates the event-time predictability in a cycle of strain-energy accumulation and release for the fore and main event (the 2011 Tohoku M9). The number of averages and the $Mc$ value may change in real-time monitoring.

The final date conversion from time $m$ becomes within a day by real-time monitoring of strain-energy accumulation and release cycles expressed by Physical Wavelets [6, 7]. The region may be near the expected epicenter area, much smaller than mesh 4°×5° while reducing the $Mc$ value. The moving sum $2s$ may be any number $w$.

Automatic detection of the anomalous peak with a sharp increase and a sudden decrease requires real-time $PW(c, j)$ monitoring in section 6 by replacing $D(c, \tau)$ with $NCI(\tau, 2w + 1)$.

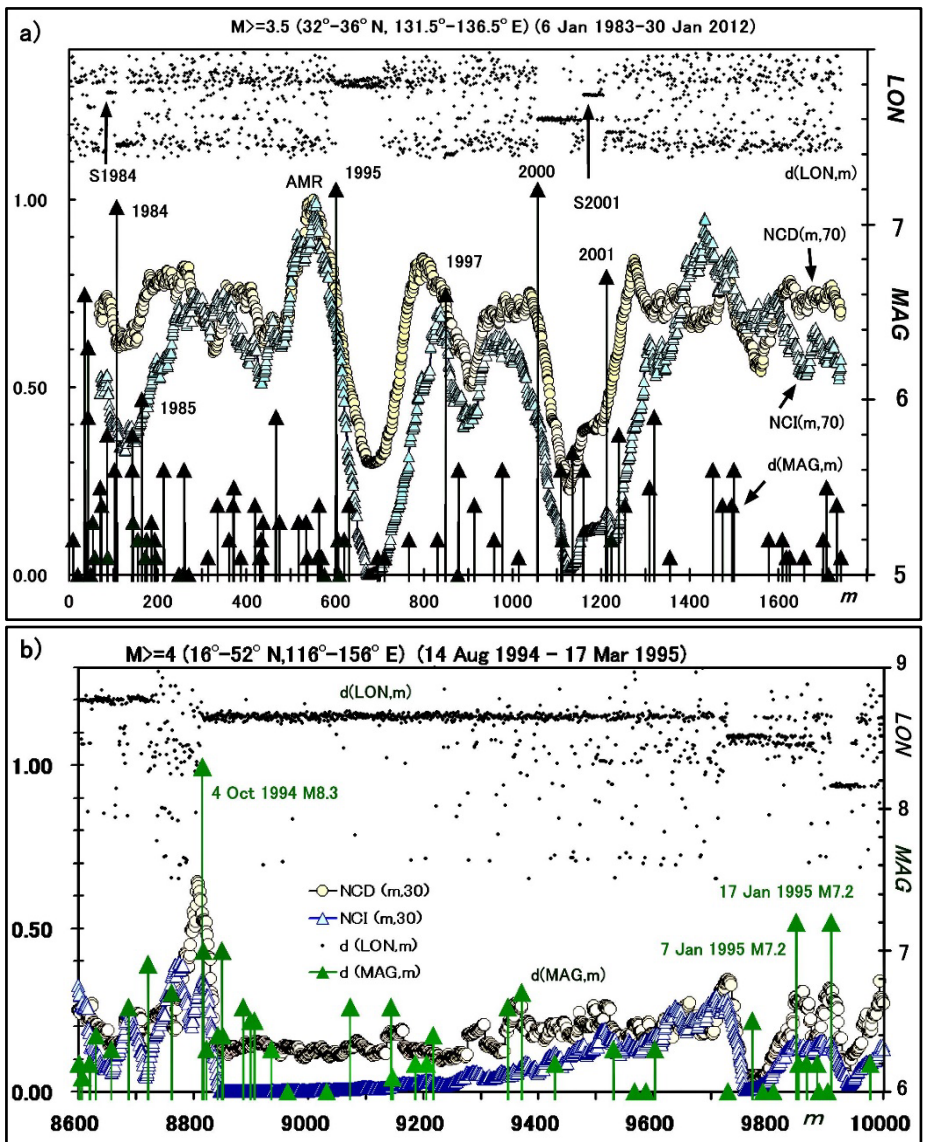

**Figure 8. Normalized strain-energy cycle on the 1995 Kobe M7.2**
Figure 8 is modified and reproduced from references [6, 7]. (a) The normalized strain-energy-density time series (cycle) $NCI(m, 70)$ and $NCD(m, 70)$ from 6 Jan 1983 to 30 Jan 2012 in the small region of $LAT = 32°–36°$ N and $LON = 131.5°–136.5°$ E as in Fig. 2a. Their time series are from shallow EQs of $MAG \geq 3.5$ and $DEP \leq 300$ km. The large EQs have labels on $d(MAG, m)$ in the full-year notation. For example, two EQ swarms are S1984 and S2001, respectively. The figure axes are the same as those in Fig.1a. A background seismicity increase, AMR, started at $NCI(m, 70)$ and $NCD(m, 70)$'s peaks and continued to the 1995 Kobe M7.2. Time $m$ has the following corresponding date: $m = 200$ to 3 Jan 1986; $m = 400$ to 28 Apr 1990; $m = 600$ to 16 Jan 1995; $m = 800$ to 26 Jun 1996; $m = 1000$ to 28 Dec 1999; $m = 1200$ to 8 Feb 2001; $m = 1400$ to 27 Oct 2004; $m = 1600$ to 15 Jul 2009. (b) The $NCI(m, 30)$ and $NCD(m, 30)$ from 14 Aug 1994 to 17 Mar 1995 in the large region of $LAT = 16°–52°$ N and $LON = 116°–156°$ E. Their time series are from all EQs of $MAG \geq 4$ from JMA focus catalogs of 1983 – 1997. The strain-energy cycles show the 4 Oct 1994 M8.3, the 7 Jan 1995 M7.2, and the 17 Jan 1995 M7.2 (Kobe event). Time $m$ has the following corresponding date: $m = 8800$ to 27 Sept 1994; $m = 9000$ to 5 Oct 1994; $m = 9200$ to 9 Oct 1994; $m = 9400$ to 21 Oct 1994; $m = 9600$ to 22 Nov 1994; $m = 9800$ to 30 Dec 1994.

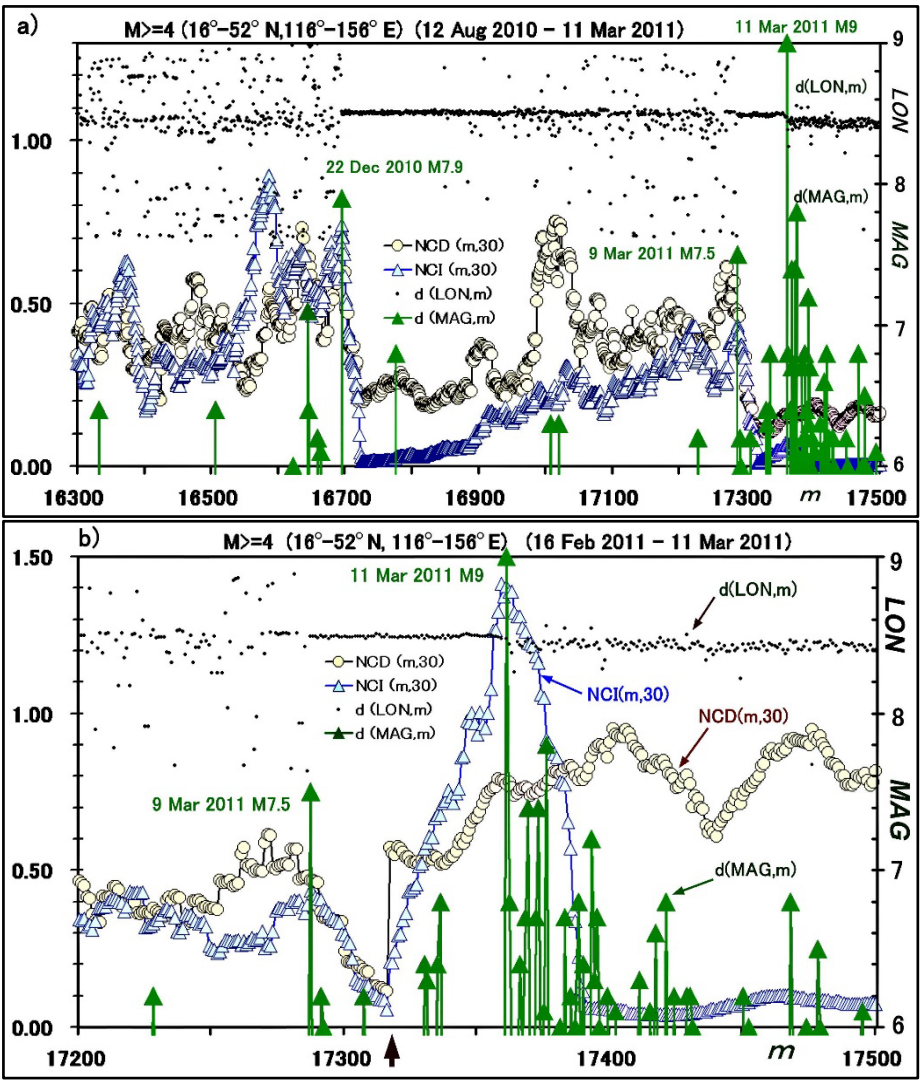

**Figure 9. Normalized strain-energy cycle on the 2011 Tohoku M9**
Figure 9 is modified and reproduced from references [6, 7]. (a) The normalized strain-energy-density time series (cycle) of $NCI(m, 30)$ and $NCD(m, 30)$ from 12 Aug 2010 to 11 Mar 2011. Their time series are for the EQs of $MAG \geq 4$ from the JMA unified focus catalogs for the region of $LAT = 16°–52°$ N and $LON = 116°–156°$ E. The figure axes are the same as those in Fig. 2 and Fig. 8. The 22 Dec 2010 M7.9 at $m = 16695$ is a precursory event to the 2011 Tohoku M9, which occurred near Chichijima (an island in the Pacific Ocean), as in Fig.3. A large foreshock to the M9 event at $m = 17287$ is the 9 Mar 2011 M7.5. Time $m$ has the following corresponding date: $m = 16500$ to 3 Oct 2010; $m = 16700$ to 22 Dec 2010; $m = 16900$ to 28 Dec 2010; $m = 17000$ to 10 Jan 2011; $m = 17100$ to 25 Jan 2011; $m = 17200$ to 16 Feb 2011; $m = 17300$ to 9 Mar 2011. (b) The normalized and magnified strain-energy cycle of $NCI(m, 30)$ and $NCD(m, 30)$ from $m = 17200$ (16 Feb 2011 at 02:23) to 17500 (11 Mar 2011 at 22:35). The $NCI(m, 30)$ and $NCD(m, 30)$ are respectively magnified by 20 and 5 times after $m = 17317$ (9 Mar 2011 at 16:56) pointed with the up-arrow on the $m$ axis. The peaks for $NCI(m, 30)$ and $NCD(m, 30)$ are at $m = 17359$ (11 Mar 2011 at 13:12) and $m = 17358$ (11 Mar 2011 at 10:41), respectively. The 11 Mar 2011 M9 (the 2011 Tohoku M9) occurred at $m = 17361$ (11 Mar 2011 at 14:46). Time $m$ has the following corresponding date and time: $m = 17300$ to 9 Mar 2011 at 13:04; $m = 17400$ to 11 Mar 2011 at 16:36.

**Concluding summary**

Seismic networks provide a chronological record of EQ events, denoted by $\{c\} = \{d(c, 1), ., d(c, m), ., d(c, N)\}$, where $c$ represents parameters such as latitude ($LAT$), longitude ($LON$), depth ($DEP$), inter-event time ($INT$), and magnitude ($MAG$), and $d(c, m)$ represents the EQ source parameter at event index $m$. However, the stochastic nature of $d(c, m)$ completely obscures subtle depth-dependent and deterministic EQ phenomena originating from the D – B transition region [7].

To resolve this issue, a physics-based EQ model has been developed [6, 7]. The model describes an EQ event as a virtual EQ particle that emerges and moves to a new location at the next event in the $c$-coordinate space. The position of the EQ particle at index time $m$ is denoted by $d(c, m)$ and a function of the principal stress components, which are full of stochastic noise. Selecting EQ events with $MAG \geq Mc$ ($Mc \approx 3.5$) reduces the noise level to about 15 ~ 25 % and enables the observation of scale-dependent EQ phenomena [6, 7, 12–15, 22–25].

However, the trajectory of the EQ particle motion is still stochastic and non-differentiable. To obtain a noise-free trajectory, Physical Wavelets can be used to describe an averaged EQ particle motion as a periodic equation of $F(c, \tau) = A(c, \tau) \approx - K(c) \times D(c, \tau)$ at the index time $\tau$. Here, $F(c, \tau)$ is the restoring force, $A(c, \tau)$ represents the acceleration, $K(c)$ is a weak positive function of time $\tau$, and $D(c, \tau)$ represents the displacement averaged over EQ particle positions, which is a noise-free function of three principal stress components in the mesh [7].

The unique trajectories approaching significant and megathrust EQ positions in the $c$-coordinate space are the EQ genesis processes named CQK and CQT. These processes have the phase inversion between $A(DEP, \tau)$ and $A(INT, \tau)$ with the negative amplitude of $A(MAG, \tau)$ weeks and months before the event ruptures. The periodic equation predicts the fault size and movement, the rupture time, and the focus of imminent significant and megathrust EQs [6, 7].

Physical Wavelets define the current monitoring time $m$ as $m = \tau + w + s$ in the real-time monitoring of CQK or CQT, as described in Eqs. (2) – (4). The CQK and CQT accompany the strain-energy accumulation and release cycles expressed with $NCI(m, 2s)$ and $NCD(m, 2s)$ in the selected mesh. These same cycles also appear in a 60 times wider mesh size surrounding the original mesh of CQK or CQT. By monitoring $NCI(m, 2s)$ and $NCD(m, 2s)$ in an appropriate mesh size, the rupture prediction event time $\tau r$ can be converted to its date and time within a day accuracy [6, 14]. Automated detection of CQK or CQT with $\tau r$ is available through $PW(m, 2s)$ monitoring.

A GPS network provides daily displacements at the stations with a time series data of $\{c\} = \{d(c, 1), ., d(c, m), .\}$, where $d(c, m)$ represents the displacement at time $m$ in days and $c$ represents geographic coordinates; $E$ (eastward), $N$ (northward), and $h$ (upward) in right-handed $(E, N, h)$. However, environmental noises in $\{c\}$ have completely masked the vital information on the crustal deformation and movement driven by tectonic and tidal forces. Physical Wavelets can quantify the underlying deformation dynamics recorded in $\{c\}$ using the $D(c, \tau) - V(c, \tau)$ and $D(c, \tau) - A(c, \tau)$ path. For example, the bulge-bending deformation on the Tohoku area (overriding tectonic plate) was a megathrust EQ genesis process of 15 months, which generated the abnormal movement of the subducting oceanic plate by coupling the two plates with the significant fault. The abnormal paths in Figs. 6 and 7 establish a real-time prediction of the megathrust EQ and tsunami generations. Observing how the oceanic plate will respond to the lunar synodic (29.5 days) tidal force loadings during the megathrust genesis processes is a key to real-time prediction and hazard mitigation measures.

The B-upper crust also responds to the periodic lunar tidal force loadings. Lunar fortnightly (14 days) force loading on the B-part expects standard unless the regional frictional failure stress state has an imminent EQ with CQK or CQT. Unusual changes from standard appear as abnormal loading in $PW(m, 2s)$ with $w \approx 2$ and $s \approx 7$ about two weeks before EQs with $MAG \geq$ about 5 [39]. Thus, along with CQK, CQT, $NCI(m, 2s)$, and $NCD(m, 2s)$, the GPS observation of lunar fortnightly and synodic tidal-force loadings with the phase-plane path and $PW(m, 2s)$ establishes a supplementary real-time prediction of the significant EQs.

Using deterministic physics-based approaches with Physical Wavelets can significantly improve the solely probabilistic approaches to significant and megathrust EQ hazard mitigation measures. Specifically, by quantifying and predicting the movements of the crust and oceanic plates during the megathrust EQ genesis processes, Physical Wavelets can offer the real-time prediction and disaster prevention warnings up to three months before the occurrence of megathrust EQ and tsunami events, saving lives and minimizing damage. These physics-based approaches demonstrate the potential of integrating innovative technology and scientific advancements to reduce the impact of natural disasters.

**Acknowledgments**

The author utilized publicly available GPS data from the Geospatial Information Authority of Japan (GSI) [5], EQ source parameters (unified hypocenter catalogs) [4], focal mechanism solutions (CMT solutions) from the Japan Meteorological Agency (JMA) [40], and a Google Earth map in this study.

The author would like to dedicate this study to the memory of four late scientists who mentored and inspired him to tackle the scientific challenges of earthquake predictions. Professor Keiiti Aki (March 3, 1930 - May 17, 2005) was a renowned geophysicist who made significant contributions to the study of earthquakes and volcanoes. Professor Makoto Takeo (April 6, 1920 - May 23, 2010) was a physicist who made significant contributions to the study of pressure broadening of atomic spectra and disperse systems and was the author's dissertation [44] advisor. Professor Gertrude Rempfer (January 30, 1912 - October 4, 2011) was a physicist who made significant contributions to the study of electron microscopes and served on the author's dissertation committee. Finally, Professor Rikiya Takeda (September 27, 1923 - March 7, 2011) was a fluid engineer who made significant contributions to the study of turbulence and flow measurement using propeller-type current meters and was the author's father.

**Appendix**

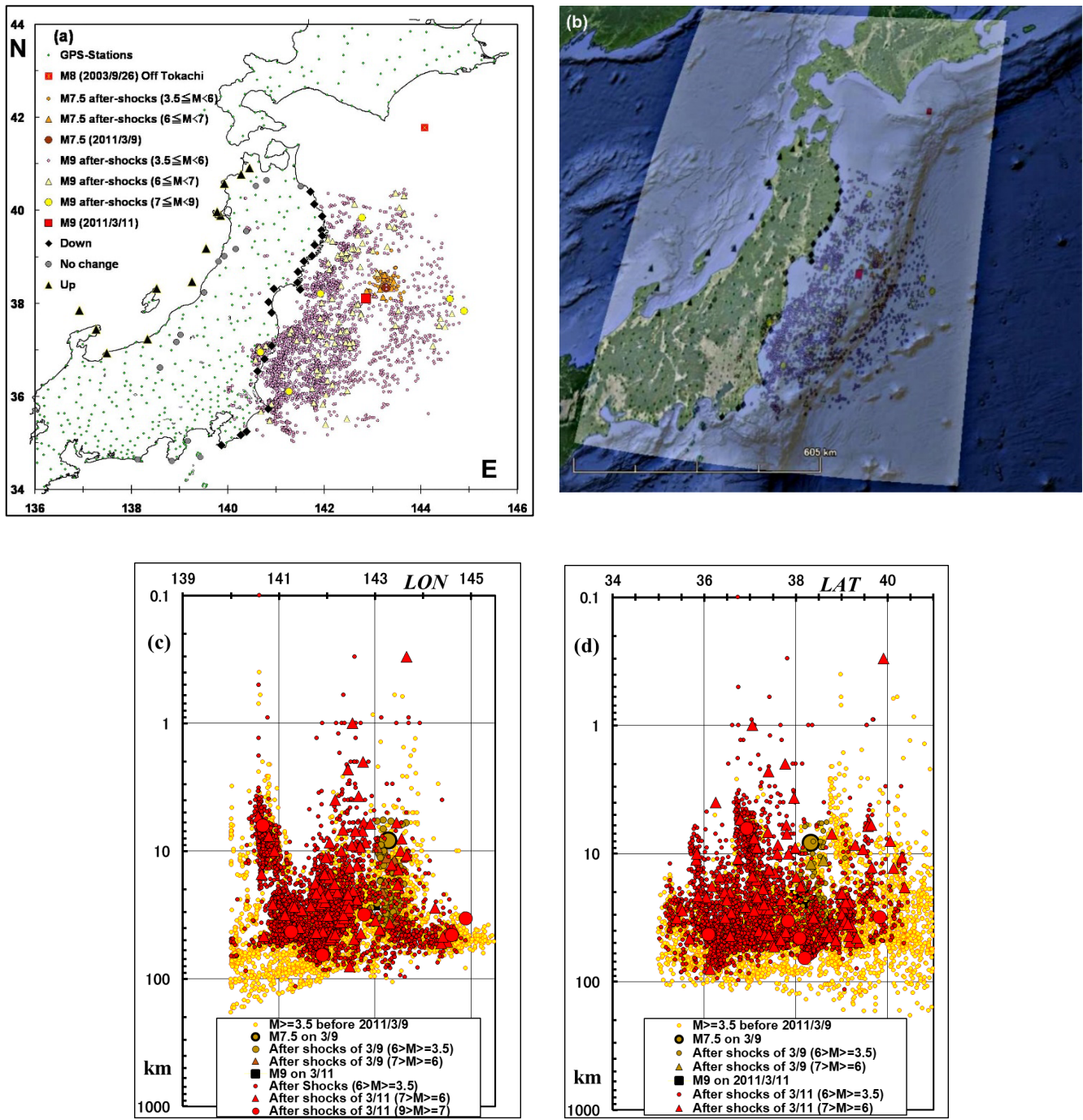

**Figure A1. Foreshocks and aftershocks of the 2011 Tohoku M9 EQ**
Figure A1 is modified and reproduced from references [6, 7]. (a) The EQ source parameters are from the JMA's unified focus catalogs. The EQ's magnitude *M* is JMA's magnitude. The M7.5 (2011/3/9) EQ is a foreshock of the Tohoku M9 EQ (2011/3/11). The M9's focus and focal mechanism were (38.1006°N, 142.8517°E, 24 km) and the reverse faulting of (STR = 193°, DIP = 10°, SLIP = 79°). Aftershocks shown in the figures are until 29 Apr 2011, and the total event number is 3634. Another significant EQ in this area was the off Tokachi M8 (2003/9/23) EQ. The vertical co-seismic displacements on 11 Mar 2011, over the 500 km distance, are the downward (Down), upward (Up), and no change (No change) displacement at each GPS station. (b) A Google Earth map with Fig. A1a overlaid. (c) The EQ focus depth, *d* (*DEP*, *m*) km, in logarithmic scale and *LON* (longitude) distribution. The EQs before M7.5 (2011/3/9) on 9 Mar 2011 are from 1 Jan 1997 to 9 Mar 2011 (one before M7.5), and the total number is 8341. (d) The depth and *LAT* (latitude) distribution.